\documentclass[11pt]{article}
\usepackage{graphicx}
\usepackage{amsmath,amssymb,amsfonts}

\newcommand{\myfig}[3]{
	\begin{figure}[h]
	\centering
	\includegraphics[width=#2cm]{#1}\caption{#3}\label{fig:#1}
	\end{figure}
	}

\newcommand{\be}{\begin{equation}}
\newcommand{\ee}{\end{equation}}
\newcommand{\bea}{\begin{eqnarray}}
\newcommand{\eea}{\end{eqnarray}}
\def\del{\partial}
\def\f{\rm{f}}
\def\D{\rm{D}}
\def\Dq{{\rm{D}}q}
\def\Dp{{\rm{D}}p}
\def\bDp{\overline{{\rm{D}}p}}

\newcommand{\p}{\partial}

\newcommand{\w}{\wedge}

\newcommand{\h}{\hspace}

\begin{document}
%\title
\rightline{hep-th/yymmnnn}
\rightline{ILL-(TH)-08-xx}
\begin{center}\ \\ \vspace{60pt}
{\Large {\bf Transversely-intersecting D-branes at finite temperature\\ and chiral phase transition}}\\ \vspace{60pt}

Mohammad Edalati, Robert G. Leigh and Nam Nguyen Hoang\\ \vspace{5pt}
{\sl Department of Physics, University of Illinois at Urbana-Champaign, Urbana IL 61801, USA}\\ \vspace{5pt}
\vspace{10pt}
\centerline{\tt edalati, rgleigh, nnguyen2@uiuc.edu, }

\end{center}
\vspace{30pt}

%\onehalfspacing

\centerline{\bf Abstract}
\noindent We consider Sakai-Sugimoto like models consisting of $\Dq$-$\Dp$-$\bDp$-branes where
$N_{\f}$ flavor $\Dp$ and $\bDp$-branes transversely intersect $N_c$ color 
$\Dq$-branes along two $(r+1)$-dimensional subspaces. For
some values of $p$ and $q$, the theory of intersections dynamically
breaks non-Abelian chiral symmetry which is holographically realized as
a smooth connection of the flavor branes at some point in the
bulk of the geometry created by $N_c$ $\Dq$-branes. We analyze the
system at finite temperature and map out different phases of the theory 
representing chiral symmetry breaking and restoration. For $q\leq 4$ we find that, 
unlike the zero-temperature case, there exist two branches of 
smoothly-connected solutions for the flavor branes, one getting very close 
to the horizon of the background and the other staying farther away from it. At 
low temperatures, the solution which stays farther away 
from the horizon determines the vacuum. For background ${\D}5$ and ${\D}6$-branes we find that the flavor branes, like the zero temperature case, show subtle behavior whose dual gauge theory interpretation is not clear. We conclude with some comments on how chiral phase transition in these models can be seen from their open string tachyon dynamics.

\newpage
\section{Introduction, summary and conclusions}

A very interesting holographic model of QCD which realizes dynamical breaking 
of non-Abelian chiral symmetry in a nice geometrical way is the
Sakai-Sugimoto model \cite{ss0412}. In this model, one starts with $N_c$
${\D}4$-branes extended in $(x^0 x^1 x^2 x^3 x^4)$-directions with $x^4$
being a circle of radius $R$. The low energy theory on  the
branes is a $(4+1)$-dimensional $SU(N_c)$ SYM with sixteen supercharges.
To break supersymmetry, anti-periodic boundary condition for fermions
around the $x^4$-circle must be chosen. To this system, $N_{\f}$ $\D 8$
and $N_{\f}$ ${\overline \D 8}$-branes are added such that they
intersect the $\D 4$-branes at two $(3+1)$-dimensional subspaces
$\mathbb{R}^{3,1}$, and are separated in the compact $x^4$-direction by
a coordinate distance of $\ell_0 = \pi R$. In other words, the $\D 8$
and $\overline \D 8$-branes are located asymptotically at the antipodal points on the
circle. The massless degrees of freedom of the system are the gauge
bosons coming from $4-4$ strings and chiral fermions coming from $4$-$8$
strings. (Before compactifying $x^4$, there are also massless adjoint
scalars and fermions coming from the $4-4$ strings. These modes become
massive upon compactifying $x^4$ and choosing anti-periodic boundary
condition for the adjoint fermions; fermions get masses at tree level
whereas scalars get masses due to loop effects.) The fermions localized
at the intersection of $\D 4$ and $\D 8$-branes have (by definition)
left-handed chirality and the fermions at the intersection of $\D 4$ and
$\overline \D 8$-branes are right-handed. The $U(N_{\f}) \times
U(N_{\f})$ gauge symmetry of the $\D 8$ and $\overline \D 8$-branes in
this model is interpreted as the chiral symmetry of the fermions living
at the intersections.  At weak effective four-dimensional 't Hooft coupling
$\lambda_4$ the low energy  theory contains QCD but this is not the
limit amenable to analysis by the gauge-gravity duality. At
strong-coupling ($\lambda_4 \gg 1$), however, the theory is not QCD but
can be analyzed using the gauge-gravity duality \cite{m9711, gkp9802, 
w9802, agmoo}, and it has been suggested that it is
in the same universality class as QCD. In lack of any rigorous proof for
this universality, the best one can do is to check whether this model at
large $N_c$ and large four-dimensional 't Hooft coupling exhibits the
key features of QCD; namely, confinement and spontaneous chiral symmetry
breaking. In fact, it apparently does \cite{ss0412}. By considering $N_{\f}$ $\D 8$
and $N_{\f}$ $\overline \D 8$-branes as probe ``flavor" branes in the
near-horizon geometry of $N_c$ ``color" $\D 4$-branes and analyzing 
the Dirac-Born-Infeld (DBI) action of the flavor branes in the background, 
one observes that at some radial point in the bulk the preferred configuration of the flavor
branes is that of smoothly-connected $\D 8$ and $\overline \D
8$-branes. This geometrical picture is interpreted as dynamical breaking
of $U(N_{\f})\times U(N_{\f})$ chiral symmetry (where the branes are
asymptotically separated) down to a single $U(N_{\f})$ (where the 
branes connect). The model also shows confinement \cite{w9803, bisy9803}.

Although choosing $\ell_0 = \pi R$ makes calculations a bit simpler,
there is no particular reason to consider the flavor branes to be
asymptotically located at the antipodal points of $x^4$. In fact, the
$\ell_0 \ll \pi R$ limit of the Sakai-Sugimoto model is very interesting
in its own right. By analyzing the $\ell_0 \ll \pi R$ limit, or
equivalently the $R\to \infty$ limit, it was realized in \cite{ahjk0604}
that the theory at the intersections can be analyzed both at weak and
strong effective four-dimensional 't Hooft coupling $\lambda_4$. At
weak-coupling the model can be analyzed using field theoretic methods
and in fact is a non-local version of Nambu-Jona-Lasinio (NJL) model \cite{njl1961}. At
strong-coupling it can be analyzed by studying the DBI action of the
flavor branes in the near-horizon geometry of the color branes and
exhibits chiral symmetry breaking via a smooth fusion of the flavor
branes at some (radial) point in the bulk. A nice feature of this model
is that the scale of chiral symmetry breaking is different from that of
confinement \cite{ahjk0604}, and one can completely turn off confinement
by taking the $R \to \infty$ limit. Therefore the $\ell_0 \ll \pi R$
limit of the Sakai-Sugimoto model provides a clean holographic model of
just chiral symmetry breaking without complications due to confinement.

The finite-temperature analysis (at large $N_c$ and large 't Hooft
coupling $\lambda_4$) of the Sakai-Sugimoto model as well as the holographic NJL model \cite{ahjk0604}, was carried out in \cite{asy0604, 
ps0604}. Putting the flavor branes as probes ($N_{\f}\ll N_c$) in the
near-horizon geometry of $N_c$ non-extremal $\D 4$-branes and analyzing
the DBI action of the flavor branes, one obtains \cite{asy0604, ps0604}
that at low temperatures (compared to $\ell_0^{-1}$) the
energetically-favorable solution is that of smoothly-connected $\D 8$
and $\overline \D 8$-branes which, like its zero-temperature
counterpart, is a realization of chiral symmetry breaking. At high
enough temperatures, on the other hand, the preferred (in the path
integral sense) configuration is that of disjoint $\D 8$ and $\overline
\D 8$-branes, hence chiral symmetry is restored. Also, when $x^4$ is
compact and $\ell_0<\pi R$, there exists an intermediate phase 
where the dual gauge theory is deconfined while chiral symmetry is
broken \cite{asy0604}.

It is certainly interesting to explore whether the holographic
realization of chiral symmetry breaking and restoration is specific to a
particular model such as the aforementioned ones, or generic in the
sense that other intersecting brane models will realize it, too. To
search for genericness (or non-genericness) of chiral symmetry breaking
in intersecting brane models, a system of $\Dq$-$\Dp$-$\bDp$-branes was
considered at zero temperature in \cite{ahk0608} where the color
$\Dq$-branes are stretched in non-compact $(x^0 x^1 \dots
x^q)$-directions. The flavor $\Dp$ and $\bDp$-branes intersect the color
branes at two $(r+1)$-dimensional subspaces $\mathbb{R}^{r,1}$. Without
flavor branes, the low energy theory on the color branes and whether it
can be decoupled from gravity (or other non-field theoretic degrees of 
freedom) with an appropriate scaling limit was analyzed in \cite{imsy9802}. 
The low energy theory on the $\Dq$-branes
is asymptotically free for $q<3$, conformal for $q=3$, and infrared free
for $q>3$. While for $q\leq5$, there always exists a scaling limit for
which the open string modes can be decoupled from the closed string
modes, there exists no such limit for $q=6$. With the flavor branes
present, an analysis was carried out in \cite{ahk0608} where the
behavior of the model for both small and large effective 't Hooft
coupling $\lambda_{\rm{eff}} \sim \lambda _{q+1} \ell_0^{3-q}$ ($\ell_0$
is the asymptotic coordinate distance between  $\Dp$ and $\bDp$-branes,
and $\lambda _{q+1}$ is the 't Hooft coupling of the $(q+1)$-dimensional
theory on the $\Dq$-branes.) was considered. Taking the 
$N_c \to \infty$ limit while keeping $\lambda_{\rm{eff}}$ fixed and large, 
amounts, in the probe approximation, to putting the flavor branes in the 
near-horizon geometry of $N_c$ color branes. (See \cite{ahk0608, imsy9802} for the 
validity of the supergravity analysis in these models.) Determining 
the shape  of the flavor branes (relevant for chiral
symmetry breaking) by analyzing their DBI action in the background
geometry, it was observed that \cite{ahk0608} for $q\leq 4$ there always
exists a smoothly-connected brane solution which is energetically
favorable. For a subclass of these general intersecting brane models, namely those for 
which $q+p-r=9$, the above-mentioned connected solutions can be
identified with the $U(N_{\f})\times U(N_{\f})$ chiral symmetry being
spontaneously broken. Following \cite{ahk0608} we will call the brane
models for which $q+p-r=9$ as transversely-intersecting brane models and
their intersections as transverse intersections. For $q=5$, there exists
no connected solution except when $\ell_0$ takes a particular value. For
this particular value of $\ell_0$ which is around the scale of non-locality 
of the low energy theory on the $\D 5$-branes (which is a little string 
theory; see \cite{a9911} for a review of little string theories), there
is a continuum of connected solutions whose turning points can be
anywhere in the radial coordinate of the bulk geometry.  All such
solutions are equally energetically favorable. For $q=6$, there is a
connected solution but is not the preferred one. The more energetically
favorable solution, in this case, is that of disjoint branes. Due to the
lacking of an appropriate decoupling limit for $q=6$, it is not clear
whether one can realize such a solution as a phase for which chiral
symmetry is unbroken. 

\paragraph{Summary and conclusions}

\noindent The purpose of this paper is to investigate some aspects of
non-compact transversely-intersecting D-brane models at finite temperature, in particular 
the number of solutions, their behavior, and whether or not such solutions can be 
identified with a chirally broken (or restored) phase of the dual gauge theory living 
at the intersections. The main reason for us to consider transverse intersections is that in the probe approximation the generalization of the Abelian $U(1)\times U(1)$ chiral symmetry to the
non-Abelian case is straightforward: one just replaces $N_{\f}=1$ with
general $N_{\f}$, and multiplies the flavor DBI action by $N_{\f}$. 
This is {\it not} the case in other holographic models of
chiral symmetry breaking and restoration\footnote{For non-transverse
intersections, namely $q+p-r \neq 9$, one can identify a symmetry in the
common transverse direction as a chiral symmetry for the fermions of the
intersections \cite{kmmw0311}. In some cases the generalization to
low-rank non-Abelian chiral symmetry is possible \cite{hnt0710}, but
such generalizations are not generic.}. Note that the Sakai-Sugimoto
model \cite{ss0412}, its non-compact version \cite{ahjk0604}, and the
$\D 2$-$\D 8$-$\overline \D 8$ model analyzed in \cite{gxz0605} which
holographically realizes a non-local version of the Gross-Neveu model 
\cite{gn1974} are examples of transverse intersections.

This paper is organized as follows. In section two we first review  the
general set up of the transversely-intersecting
$\Dq$-$\Dp$-$\bDp$-branes, then consider the system at finite
temperature at large $N_c$ and large effective 't Hooft coupling
$\lambda_{\rm eff}$. There are two saddle point contributions (thermal and black brane) to the bulk Euclidean  path integral, and each can potentially be used as a background geometry dual to the color theory. Comparing the free energies of the two saddle points, we determine which one is dominant (has lower free energy) for various $q$'s. For $q\neq5$, the dominant saddle point is the black brane geometry, whereas for $q=5$, the thermal geometry is typically dominant.  In section three we present the solutions to the equation of motion for the DBI action of the flavor branes placed in the near horizon geometry of black $\Dq$-branes. By a combination of analytical and numerical techniques, we show that, unlike the zero-temperature
case, for $\ell_0 /\beta $ less than a critical value, there exist
generically two branches of smoothly-connected solutions (and of course,
a solution with disjoint branes) for $q\leq 4$. ($\beta$ is the circumference of the asymptotic Euclidean time circle, and is equal to the inverse of the dual gauge theory temperature $T$.) 
Note that some of these branches were previously missed in the literature. One branch which we will call the ``long" connected solution gets very close to the horizon of the black $\Dq$-branes whereas the other one named ``short" connected
solution stays farther away from it. Beyond the critical value, the
flavor branes are ``screened" and cannot exist as a
connected solution. The situation is, however, different for $q=5,6$.
For $q=5$ and $T<(2\pi R_{5+1})^{-1}$, where $R_{5+1}$ denotes the characteristic radius of the $\D5$-brane geometry, the flavor branes must be placed in the near horizon geometry of thermal $\D5$-branes.  Like the zero temperature case, we find that there exists an infinite number of connected solutions when $\ell_0$ is around the non-locality scale  of the low energy theory on the $\D5$-branes. This scale is set by the (inverse) hagedorn temperature of $\D5$-brane little string theory. There are no connected solution for other values of $\ell_0$, though. For $T=(2\pi R_{5+1})^{-1}$ the flavor branes should be considered in the near horizon geometry of black $\D5$-branes. In this case there always exists one connected 
solution, as well as a disjoint solution, for small $\ell_0$ (compared to $(2\pi R_{5+1})^{-1}$). 
In fact, at this particular temperature there is one connected solution for $\ell_0$'s much less than $R_{5+1}$. The number of solutions for $\ell_0$ beyond the non-locality scale ($\sim R_{5+1}$)  depends on the dimension of the intersections. For  four-dimensional intersections, there 
are two connected solutions up to a critical value whereas for 
two-dimensional intersections there is no connected solution. Of course, for both two- and four-dimensional intersection there is always a solution representing disjoint branes. Having determined the flavor brane solutions in the background of thermal and black $\D5$-branes,  it  is not clear whether or how these solutions represent a chirally-broken or restored phase in the dual theory, mainly because in the geometry of $\D5-branes$, there are modes (non-field theoretic)  which cannot totally be decoupled from the dual field theory degrees of freedom. Lastly, for $q=6$, independent of what value $\ell_0 /\beta$ takes, there always exist one connected solution (and a disjoint solution). 

In section four, we map out different phases of the dual gauge theories and determine whether or not there is a chiral symmetry breaking-restoration phase transition. We do this by
comparing the regularized free energies of the various branches of the solutions found in
section three. For $q\leq 4$ the short solution is preferred to both
long and disjoint solutions at small enough temperatures (compared to
$\ell_0^{-1}$), hence chiral symmetry is broken. At high temperatures,
however, chiral symmetry gets restored and this phase transition is 
first order. For $q=5$ and $T<(2\pi R_{5+1})^{-1}$, the infinite number of connected solutions are all equally energetically favorable, and each one of them is preferred over the disjoint solution. For $T=(2\pi R_{5+1})^{-1}$, we find that for small enough ${\ell_0}/(2\pi R_{5+1})$ the disjoint solution is preferred. For larger values of $\ell_0$, there
is no connected solution so the disjoint solution is the vacuum. As we alluded to earlier, it is not clear to us that the preferred solutions of the flavor branes in the background of color $\D5$-branes can be associated with different phases of the dual theory. For $q=6$, although we find that the disjoint solution is always preferred and there is no phase transition, there is no clear way to associate this solution with unbroken chiral symmetry in the dual field theory. This is because the is no decoupling limit that one can take to separate the gravitational degrees of freedom of those of the dual theory.  

Section five is devoted to a brief analysis of the number of solutions and 
their energies for transversely-intersecting $\Dq$-$\Dp$-$\bDp$-branes 
with compact $x^q$. In section six we speculate how the order 
parameter for chiral symmetry breaking can be realized in finite-temperature 
transversely-intersecting D-branes by including the thermal 
dynamics of an open string tachyon stretched between the flavor branes, 
and how it may depend on temperature. Finally, in the appendix we present detailed calculations  for the free energies of the near horizon geometries of color $\Dq$-branes (with the topology of  either ${\rm S}^1\times {\rm S}^1$ or ${\rm S}^1\times \mathbb{R}$ in the $t-x^q$ submanifold) to determine the dominant background geometry (either thermal or black brane) at low and high temperatures.

\section{Transverse intersections at finite temperature}
We start this section by reviewing first the general setup for 
transverse  intersections  of $\Dq$-$\Dp$-$\bDp$-branes and identifying the
massless degrees of freedom at intersections. We consider the
system at finite temperature in the large $N_c$ and large 't Hooft
coupling limits. Since there is more than one background, we 
determine the one with the lowest free energy and consider that as the background dual to the color sector of the dual theory at finite temperature. 
We then write the equation of motion for the flavor branes.
This section is followed in the next section by an analysis of the
solutions to the equation of motion as a function of the dimensions of
the intersections $r$, as well as $q$. (For transverse
intersections $p$, the spacial dimension of the flavor branes, is
determined once $r$ and $q$ are given.)

\subsection{General setup}
Consider a system of intersecting $\Dq$-$\Dp$-$\bDp$-branes in flat non-compact ten-dimensional Minkowski space where $N_c$ $\Dq$-branes are stretched in $(x^0 x^1 \dots x^q)$-directions, and each stack of  $N_{\f}$ $\Dp$ and $N_{\f}$ $\bDp$-branes are extended in $(x^0 x^1 \dots
x^r)$- and $(x^{q+1} \dots x^9)$-directions. The $\Dp$- and 
$\bDp$-branes are separated in the $x^q$-direction
by a coordinate distance $\ell_0$ and intersect the $\Dq$-branes at two  $(r+1)$-dimensional intersections
\be
\begin{array}{c@{\hspace{.1in}\extracolsep{.2in}} cccccccl}
 & x^0 &  x^1 & \dots & x^r & \dots & x^q & \dots\dots & x^9 \\
\Dq & \times & \times & \dots & \times & \dots & \times & \dots\dots & .\\
\Dp & \times & \times & \dots & \times & \dots & . & \dots\dots & \times\\
\bDp & \times & \times & \dots & \times & \dots & . & \dots\dots & \times.
\end{array}
\ee
Using T-duality one can determine the massless degrees of freedom localized 
at the intersection. It turns out that for transverse intersections, $q+p-r=9$, the massless
modes which come from the Ramond sector in the $p-q$ strings are Weyl
fermions. These fermions are in the fundamentals of $U(N_c)$ and
$U(N_{\f})$. The massless modes at the other intersection are also Weyl
fermions which  transform in the fundamentals of $U(N_c)$  and 
$U(N_{\f})$ of the $\bDp$-branes.

One way to put the above system at finite temperature (at large $N_c$, large
effective 't Hooft coupling $\lambda_{\rm eff}$, and in the probe
approximation $N_{\f} \ll N_c$) is to start with the geometry of black $\Dq$-branes as background. The Euclidean metric for this geometry is
\bea\label{Dq metric}
ds^2=\left(\frac{u}{R_{q+1}}\right)^{\frac{7-q}{2}}\Big(f(u) dt^2 + d{\vec x} ^2
\Big) + \left(\frac{u}{R_{q+1}}\right)^{-\frac{7-q}{2}}
\Big(\frac {du^2}{f(u)}+ u^2 d{\Omega^2_{8-q}}\Big),
\eea
with
\bea\label{warpfactor}
f(u)= 1-\Big(\frac{u_T}{u}\Big)^{7-q},
\eea
where in the metric ${d\Omega^2_{8-q}}$ is the line element of a
$(8-q)$-sphere with a radius equal to unity, and $R_{q+1}$, which denotes the 
characteristic radius of the geometry, is given by
\bea\label{radius}
R_{q+1}^{7-q}= (2\sqrt{\pi})^{5-q} \Gamma\Big( \frac {7-q}{2}\Big) g_s  N_c l_s^{7-q}=
2^{7-2q} (\sqrt{\pi})^{9-3q} \Gamma\Big( \frac {7-q}{2}\Big) g_{q+1}^2 N_c l_s^{10-2q},
\eea
where $g_s$ is the string coupling. The Euclidean time is periodically identified: $t \sim t +\beta$, where $\beta$ is equal to the inverse of the temperature $T$ of the black branes. In (\ref{warpfactor}) the horizon radius $u_T$ is related to $\beta$ as
\bea
T=\beta^{-1}=\frac{7-q}{4\pi}\Big(\frac{u_T}{R_{q+1}}\Big)^{\frac{7-q}{2}}\frac{1}{u_T}.
\eea
The relationship between $u_T$ 
and $\beta$ comes about in order to avoid a conical singularity in 
the metric at $u=u_T$. Note that for black $\D5$-branes $\beta$ is independent of $u_T$, and equals $2\pi R_{5+1}$. 

Also, the dilaton $\phi$ and the $q$-form RR-flux $F_q$ are given by
\bea\label{dilatonq-form}
e^\phi=g_s \left(\frac{u}{R_{q+1}}\right)^{\frac{1}{4} (q-3)(7-q)},
\qquad F_q = \frac{2\pi N_c}{V_{8-q}} \epsilon_{8-q},
\eea
where $V_{8-q}$ and $\epsilon_{8-q}$ are the volume and the volume form
of the unit $(8-q)$-sphere, respectively.

There is, however, another background with the same asymptotics as 
(\ref{Dq metric}) which may potentially compete with the aforementioned
background. The metric for this (thermal) geometry is
\bea\label{Dqthermal}
ds^2=\left(\frac{u}{R_{q+1}}\right)^{\frac{7-q}{2}}\Big(dt^2 + d{\vec x} ^2
\Big) + \left(\frac{u}{R_{q+1}}\right)^{-\frac{7-q}{2}}
\Big(du^2+ u^2 d{\Omega^2_{8-q}}\Big),
\eea
with the Euclidean time $t$ being periodically identified with a period $\beta=T^{-1}$.  Unlike the black brane geometries (\ref{Dq metric}), $\beta$ could take arbitrary values in the thermal geometries (\ref{Dqthermal}). The dilaton and  the $q$-form RR-flux are the same as (\ref{dilatonq-form}).

In the appendix we have calculated the free energies of both thermal and black brane geometries. Except for $q=5$, the difference in free energies  $\Delta S$ of  the two geometries subject to the same asymptotics is given by
\bea\label{difffree}
\Delta S= S_{\rm thermal}-S_{\rm black~brane}=\frac{9-q}{g_s^2} V_9u_T^{7-q},
\eea
where $V_9$ is the volume of space transverse to the radial coordinate $u$: $V_9={\rm Vol}({\rm S}^{8-q}){\rm Vol}(\mathbb{R}^{q}){\rm Vol}({\rm S}^1_{\beta})$. The volume $V_9$ is measured in string unit $l_s $ where for simplicity we set $l_s=1$. The difference in free energies (\ref{difffree}) shows that the thermal background is less energetically favorable compared to the the black brane background (\ref{Dq metric}). Thus, there is no Hawking-Page type transition between the two geometries which holographically indicates that there is no confinement-deconfinement phase transition in the dual theory. The situation is different for $q=5$. Semi-classically, once the characteristic radius $R_{5+1}$ is given, the black $\D5$-brane geometry will have a fixed temperature $T=(2\pi R_{5+1})^{-1}$. At this specific temperature, there are two saddle points contributing to the (type IIB) supergravity path integral:  thermal and black $\D5$-brane geometries. The difference in free energies of the two saddle points is (see the appendix for more details) 
\bea
S_{\rm thermal~\D5}-S_{\rm black~\D5-brane}=\frac{8\pi}{g_s^2} {\rm Vol}({\rm S}^{3}){\rm Vol}(\mathbb{R}^{5}) R_{5+1} u_T^2, \qquad {\rm at} \qquad \beta=2\pi R_{5+1},
\eea
showing that the black brane geometry is the saddle point with lower free energy. However, at temperatures other than $(2\pi R_{5+1})^{-1}$, the thermal geometry is the only saddle point, although due to the hagedorn temperature of the $\D5$-brane theory, one should only consider temperatures less than $(2\pi R_{5+1})^{-1}$. Thus, for $T<(2\pi R_{5+1})^{-1}$ we use the thermal $\D5$-brane geometry as background.

\subsection{Flavor $\Dp$-$\bDp$-branes in black $\Dq$-brane geometries}
We are interested in the dynamics of the flavor $\Dp$ and
${\bDp}$-branes in the background of the black $\Dq$-brane geometries. As we alluded to earlier, for $q=5$ the thermal geometry of $N_c$ $\D5$-branes (once the limit of the near horizon geometry is taken) is the background that one should use for the dual finite temperature field theory for $T<(2\pi R_{5+1})^{-1}$. We also analyze the dynamics of the flavor branes in the background of the black $\D5$-branes in which case it is understood that the dual theory is at a fixed temperature $T=(2\pi R_{5+1})^{-1}$.
We are interested in the static shape of the flavor
branes as a function of the radial coordinate $u$. Therefore we choose
the embedding
\be\label{embedding}
\begin{array}{c@{\hspace{.1in}\extracolsep{.2in}} cccccl}
t=\sigma^0, &  x^1=\sigma^1, & \dots & x^r=\sigma^r, & x^q=\sigma^q,\\
u=u(\sigma^q), & x^{q+2}=\sigma^{q+1}, &  \dots & x^8=\sigma^{p-1}, & x^9=\sigma^p,
\end{array}
\ee
subject to the boundary condition
\bea\label{Dbranebc1}
u(\pm \frac{\ell_0}{2})=\infty,
\eea
where $\{\sigma^0, \cdots, \sigma^p\}$ are the worldvolume coordinates of the flavor branes. This boundary condition simply states that the asymptotic coordinate
distance between the $\Dp$ and $\bDp$-branes is $\ell_0$. Ultimately 
the stability of such an assumption lies in the large $N_c$ limit.

From now on, we set $N_{\f}=1$. As it becomes apparent in what follows, 
the generalization to $N_{\f} \ll N_c$ is straightforward. The dynamics of a $\Dp$-brane (and
a $\bDp$-brane) is determined by its DBI plus Chern-Simons action.
Solving the equations of motion for the gauge field, one can safely
set the gauge field equal to zero and just work with the DBI part of the
action. After all, it is this part of the full action which is relevant
for our purpose of determining the shape of the $\Dp$ and $\bDp$-branes.
Therefore, with gauge field(s) set equal to zero, the dynamics is captured by the DBI action
\bea\label{DBIaction}
S_{\rm{DBI}}=\mu_p \int d^{p+1}\sigma~e^{-\phi} \sqrt{\hbox{det}(g_{ab})},
\eea
where $\mu_p$ is a constant, and $g_{ab}=G_{MN}\del_a x^M \del_b x^N$ is
the induced metric on the worldvolume of the $\Dp$-brane. For a
$\Dp$-brane forming a curve $u=u(x^q)$, the DBI action (\ref{DBIaction})
reads
\bea\label{Dactionemb1}
S_{\rm{DBI}}= \beta\,C(q,r)\int d^rx~dx^q ~ u^{\frac{\gamma}{2}} \Big[f(u)+\Big(\frac{u}{R_{q+1}}\Big)^{2\delta}{u^{'}}^2\Big]^{\frac{1}{2}},
\eea
where
\bea\label{parameters}
C(q,r)&=& \frac{\mu_p}{g_s}~\hbox{Vol}({\rm S}^{8-q}) {R_{q+1}}^{\frac{1}{4} (q-7)(r-3)},\nonumber\\
\gamma &=& 2+ \frac{1}{2}(7-q)(r+1),\\
\delta &=& \frac{1}{2} (q-7)\nonumber,
\eea
and $u^{'}= du/dx^q$. For a $\Dp$-brane forming a curve $u(x^q)$ in the thermal $\D5$-brane geometry the DBI action is obtained by setting $f(u)=1$ in (\ref{Dactionemb1}). The integrand in (\ref{Dactionemb1}) does not explicitly depend on
$x^q$, therefore ${\cal L}-u^{'}\del {\cal L}/\del u^{'}$ must be
conserved (with respect to $x^q$). A first integral of the equation of motion is then obtained
\bea\label{braneeom}
{u^{\frac{\gamma}{2}} f(u)}\Big[f(u)+\Big(\frac{u}{R_{q+1}}\Big)^{2\delta}{u^{'}}^2\Big]^{-\frac{1}{2}}=u_0^{\frac{\gamma}{2}},
\eea
where $u_0$ parametrizes the solutions. We now analyze the solutions of (\ref{braneeom}).

\section{Multiple branches of solutions}

The simplest solution of the equation of motion in (\ref{braneeom}),
namely $u_0=0$, corresponds to $x^q=\hbox{constant}$. In order to
satisfy the boundary condition (\ref{Dbranebc1}), one obtains $x^q=\pm
{\ell_0}/2$. So, the $u_0=0$ solution corresponds to disjoint $\Dp$ and
$\bDp$-branes descending all the way down to the horizon at $u=u_T$. Also, note that the existence of this solution is independent of $\beta$.

For $u_0\neq 0$, solving for $u^{'}$ yields
\bea\label{uprime}
{u^{'}}^2=\frac{1}{{u_0}^\gamma} \Big(\frac{u}{R_{q+1}}\Big)^{-2\delta} f(u)~\Big(u^\gamma f(u) -{u_0}^\gamma\Big).
\eea
Since the left hand side of (\ref{uprime}) is non-negative, the right
hand side of (\ref{uprime}) must also be non-negative resulting in
$u\geq\hbox{max}\{u_T, u_{*}\}$, where $u_{*}$ is a possible turning
point. Therefore, for allowed solutions one must have $u\geq
u_{*}>u_T$.

The possible turning points are determined  by analyzing the zeros of
the right hand side of (\ref{uprime}). Setting $f(u_{*})=0$ will not
result in a  valid turning point. So, the other possibilities 
come from solving $u_{*}^\gamma f(u_{*}) -u_0^\gamma =0$, which we 
will rewrite as follows
\bea\label{turningpoint}
u_{*}^\gamma-u_{*}^\sigma u_T^{-2\delta}-u_0^\gamma =0,
\eea
where 
\bea
\sigma=\gamma+2\delta=2+\frac{1}{2} (7-q)(r-1).
\eea 
Note that since $r\neq 0$ (and in fact, for the cases of interest, it is
either $1$ or $3$), $\sigma$ is always a positive integer which,
combined with the fact that $\delta<0$, implies that $\sigma<\gamma$.

We use (\ref{uprime}) to relate the integration constant $u_{*}$, or
equivalently $u_0$, to the parameters of the theory, namely the
(inverse) temperature $\beta$  and the asymptotic distance between the
$\Dp$ and ${\bDp}$-branes $\ell_0$. First, rearrange (\ref{uprime}) to
get
\bea\label{xvsubrane}
x^q(u)&=&R^{-\delta} u_0^{\frac{\gamma}{2}}\int_{u_{*}}^u \Big(u^{-2\delta}-u_T^{-2\delta}\Big)^{-\frac{1}{2}}
\Big(u^\gamma-u^\sigma {u_T}^{-2\delta}-u_0^\gamma\Big)^{-\frac{1}{2}} du \nonumber\\
&=& R^{-\delta} (u_{*}^\gamma-u_{*}^\sigma u_T^{-2\delta})^{\frac{1}{2}} \int_{u_{*}}^u \Big(u^{-2\delta}-u_T^{-2\delta}\Big)^{-\frac{1}{2}}\times\\
&&\Big(u^\gamma-u^\sigma {u_T}^{-2\delta}-(u_{*}^\gamma-u_{*}^\sigma u_{T}^{-2\delta})\Big)^{-\frac{1}{2}} du,\nonumber
\eea
where in the second line we used (\ref{turningpoint}) to trade $u_0$ for
$u_{*}$. Changing to a new (dimensionless) variable $z=u/u_T$,
(\ref{xvsubrane}) becomes
\bea\label{xvszbrane}
x^q(z)=-\frac{\delta}{2\pi} \beta (z_{*}^\gamma-z_{*}^\sigma)^{\frac{1}{2}}\int_{z_{*}}^z \Big({\tilde z}^{-2\delta}-1\Big)^{-\frac{1}{2}}
\Big({\tilde z}^\gamma-{\tilde z}^\sigma -(z_{*}^\gamma-z_{*}^\sigma)\Big)^{-\frac{1}{2}} d{\tilde z},
\eea
where $z_{*}\in (1,\infty)$. Taking the $z\to \infty$ limit, we can
relate $z_{*}$ to $\beta$ and $\ell_0$
\bea\label{Lbetau0eq}
\frac{\ell_0}{\beta}=-\frac{\delta}{\pi}~(z_{*}^\gamma-z_{*}^\sigma)^{\frac{1}{2}}\int_{z_{*}}^{\infty} \Big({z}^{-2\delta}-1\Big)^{-\frac{1}{2}}
\Big({z}^\gamma-{z}^\sigma -(z_{*}^\gamma-z_{*}^\sigma)\Big)^{-\frac{1}{2}} dz.
\eea
As we mentioned earlier, the solutions to the equation of motion are
parametrized by possible value(s) of the turning point $z_*$. For a
fixed $\ell_0/\beta$, it is the number of $z_*$ which determines the
number of (connected) solutions. Thus, one has to analyze $\ell_0/\beta$
as a function of $z_*$ to determine the number of solutions for a fixed $\ell_0/\beta$.

\subsection{Analytical analysis}
There are regions of $z_*$ for which $\ell_0/\beta$ as a function of
$z_*$ can be given analytically. These are the $z_* \to 1^+$ and $z_*
\gg 1$ regions. For any $z_*$, the integral in  (\ref{Lbetau0eq}) can be
evaluated numerically. The numerical results will be presented shortly
after the analytical analysis for the two limiting cases is given.

\myfig{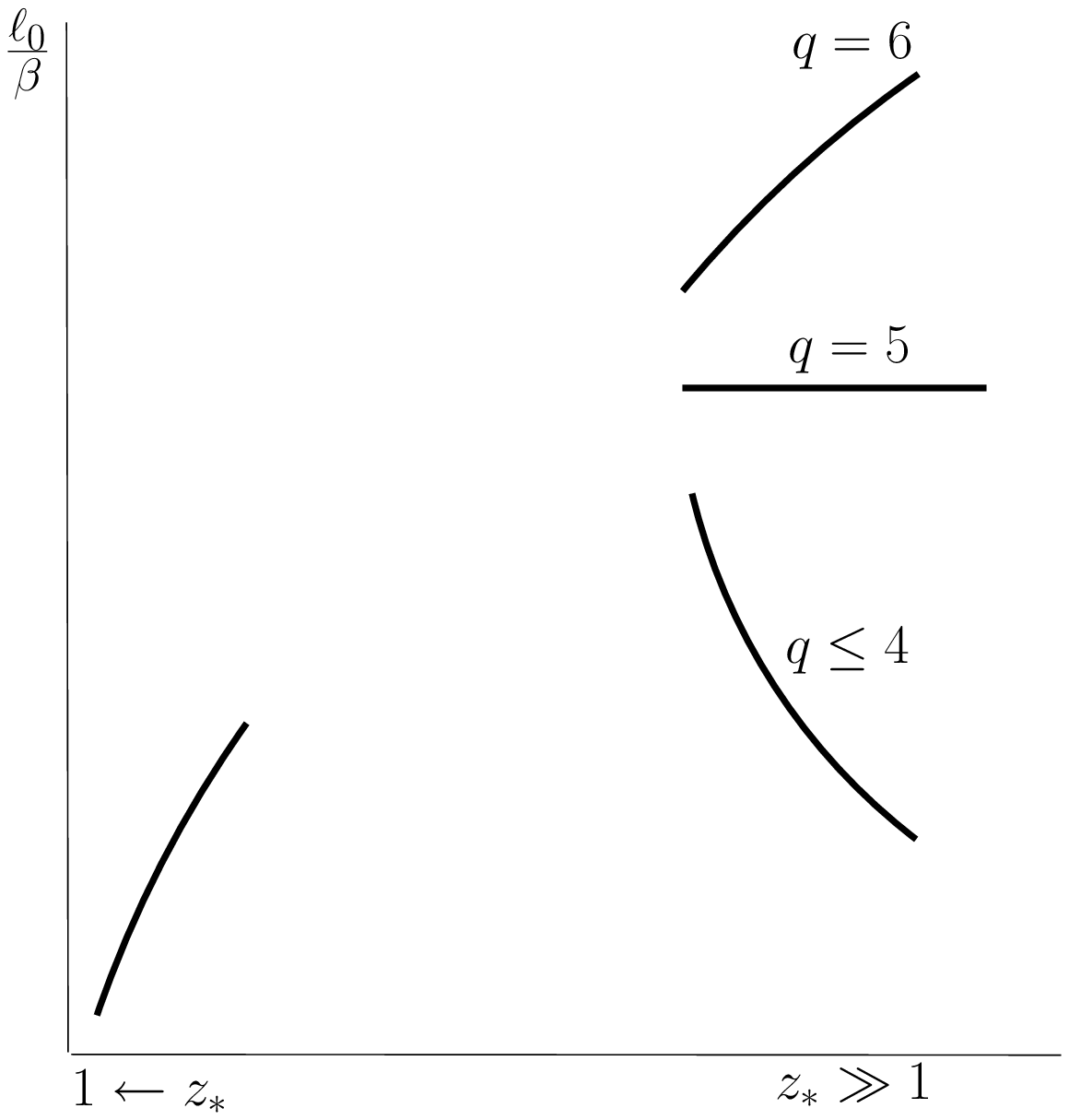}{5.2}{\footnotesize{Behavior of $\ell_0/\beta$ versus $z_*$ in two regions  of $z_* \to 1$ and $z_* \gg 1 $ for various $q$'s. Except for $q=5$, these plots illustrate the number of connected $\Dp$-branes at low and high temperatures placed in the background of black $\Dq$-branes. For $q=5$, it is understood that $\beta$ is fixed; $\beta= 2\pi R_{5+1}$, and different solutions is obtained by varying $\ell_0$.}}

First consider the $z_* \to 1^+$ limit. Taking $z_*=1+\epsilon$ where 
$0<\epsilon \ll 1$, (\ref{Lbetau0eq}) is approximated by
\bea\label{smallzstar-1}
\frac{\ell_0}{\beta}&\sim&-\frac{\delta}{\pi} \sqrt{-2\delta \epsilon}
\int_{1+\epsilon}^{\infty} (z^{-2\delta}-1)^{-\frac{1}{2}}
(z^\gamma-z^\sigma+ 2\delta\epsilon)^{-\frac{1}{2}} dz.
\eea
Ignoring some numerical prefactors, the behavior of (\ref{smallzstar-1}), is approximated by
$\sqrt{\epsilon}\int^\infty z^{\delta-\frac{\gamma}{2}} dz$ for large values of $z$. Since
$\delta-\frac{\gamma}{2} < -2$ for all $q$ and $r$ of interest, one has
$\sqrt{\epsilon}\int^\infty z^{\delta-\frac{\gamma}{2}} dz \sim
\sqrt{\epsilon}$. On the other hand, when $z $ approaches $z_*$ such that $z-z_*>0$, we define
$z=z_*+x$ with $0<x\ll 1$, and expand out the integrand of
(\ref{smallzstar-1}) around $x$.  We get
\bea\label{smallzstar-2}
\frac{\ell_0}{\beta}&\sim&-\frac{\delta}{\pi\sqrt{-2\delta}} \sqrt{\epsilon}\int_0
(x(x+\epsilon))^{-\frac{1}{2}} dx\nonumber\\
&\sim& -\frac{1}{\pi} \sqrt{-2\delta} \sqrt{\epsilon}
\log{\sqrt{\epsilon}},
\eea
indicating that the leading behavior of $\ell_0/\beta$ in the $z_* \to
1$ limit is $-\sqrt{\epsilon} \log{\sqrt{\epsilon}}$.

For the $z_* \gg 1$ region, we can approximate (\ref{Lbetau0eq}) by 
(recall $\delta <0$ and $\sigma<\gamma$)
\bea\label{largezstar}
\frac{\ell_0}{\beta}&\sim&-\frac{\delta}{\pi}~z_{*}^{\frac{\gamma}{2}}\int_{z_{*}}^{\infty}
{z}^{\delta}
\Big({z}^\gamma -z_{*}^\gamma\Big)^{-\frac{1}{2}} dz +\cdots\nonumber\\
&=& - \frac{\delta}{\pi}z_{*}^{1+\delta} \int_{1}^{\infty} {y}^{\delta}
\Big({y}^\gamma -1\Big)^{-\frac{1}{2}} dy+\cdots\nonumber\\
&=& - \frac{\delta}{\gamma\sqrt{\pi}}\frac{\Gamma\Big[\frac{\gamma-2(1+\delta)}{2\gamma}\Big]}
{\Gamma\Big[\frac{\gamma-(1+\delta)}{\gamma}\Big]}~ z_{*}^{1+\delta}+\cdots,
\eea
where $\cdots$  represents terms subleading in $z_*$, and in the second line in (\ref{largezstar}) 
we have changed the variable from $z$ to $y=z/z_*$. Thus, aside from a numerical factor, the
$z_* \to \infty$ limit of (\ref{Lbetau0eq}) reads
\bea\label{largezstar1}
\frac{\ell_0}{\beta} \sim z_{*}^{1+\delta}.
\eea
This expression is identical to the one derived for the zero temperature
case in \cite{ahk0608}. This resemblance is not accidental because the
large $z_*$ limit corresponds to having a turning point very far away
from the horizon of the background geometry. The results obtained for
this limit should then match those derived for the zero temperature
case. An interesting feature of (\ref{largezstar1}) is that for the black
$\D 5$-branes $\ell_0/(2\pi R_{5+1})$ is independent of $z_*$ and
approaches a constant value of $1/(r+3)$. This value has been argued in
\cite{ahk0608} to be around the scale of  non-locality of the low
energy effective theory on $\D5$-branes. The analysis for the dynamics of the flavor branes placed in the thermal $\D5$-brane geometry is the same as the analysis when they are placed in the zero temperature  $\D5$-brane geometry. The zero temperature analysis has already been done in \cite{ahk0608} where it was found that there exist an infinite number of connected solutions for one specific value of $\ell_0= 2\pi R_{5+1}/(r+1)$, and none for other $\ell_0$'s. 

Analyzing (\ref{Lbetau0eq}) for the two regions of $z_* \to 1^+$ and
$z_* \gg 1$, the minimum crude conclusion that one can draw is that for
small enough $\ell_0/\beta$ there exist two
connected solutions (one closer to the horizon which we will name "long"
connected solution, and the other farther away from it named "short"
connected solution) for $q\leq 4$ and only one curved solution for
$q=6$. For $q=5$ and for $T<(2\pi R_{5+1})^{-1}$, there exists an infinite number of connected solutions for just $\ell_0= 2\pi R_{5+1}/(r+1)$ and none for others. On the other hand, for $T=(2\pi R_{5+1})^{-1}$, there is only one connected solution given that $\ell_0\ll 2\pi R_{5+1}$. The analysis for the two $z_{*}$ regions has been summarized in Figure \ref{fig:pic1}.

\subsection{Numerical analysis for the number of solutions}

For generic values of $\ell_0/\beta$, the allowed number of solutions
can be determined by numerically integrating (\ref{Lbetau0eq}) and
plotting $\ell_0/\beta$ versus $z_*$. The results for various
intersections and $\Dq$-branes are shown in Figure 2. Although we have
plotted $\ell_0/\beta$ versus $z_*$ for $z_* \in (1,50)$, the
qualitative behavior of the plots stays the same if one considered
larger values of $z_*$. As we will describe below, the number of curved
solutions depends on what configuration is being considered.
Note that regardless of the configuration, there always exists a disjoint
solution. In what follows in the  rest of this subsection, when we say
there exist one or two solutions for a particular system we have
connected solutions in mind.

\paragraph{$\bold{(3+1)}$-dimensional intersections}
In this case there are three allowed D-brane configurations which are transversely 
intersecting. These are  the $\D 4$-$\D 8$-$\overline \D 8$, $\D 5$-$\D
7$-$\overline \D 7$ and $\D 6$-$\D 6$-$\overline \D 6$ configurations
shown on the top row in Figure \ref{fig:pic2}.

For the $\D 4$-$\D 8$-$\overline \D 8$ configuration (shown on the upper
left corner in Figure 2), there is a critical value of
$(\ell_0/\beta)_{\rm{cr}} \approx 0.17$ beyond which there exists no
connected solution for the flavor branes. Below this critical value
there are two connected solutions which we earlier called the
``short" and the ``long" connected solutions. The existence of these two
types of solutions and the  critical value of 0.17 were already noted by 
the authors of \cite{ps0604} in the their analysis of the holographic NJL model at finite 
temperature. We will see in the next section that the short solution is always more
energetically favorable to the long one. Since there also exists a
disjoint solution, determining the chirally-broken or chirally-symmetric phase of the dual field theory (which is a non-local version of the NJL model \cite{ahjk0604}) is just a matter of comparing  the free energies of the disjoint and short connected solution. This will be
done in the next section.

The analysis for the $\D 5$-$\D 7$-$\overline \D 7$ and $\D 6$-$\D
6$-$\overline \D 6$ cases are, however, more subtle, and the holographic
interpretation of the solutions is less transparent for reasons to be mentioned below. 
For the $\D 5$-$\D7$-$\overline \D 7$ configuration at temperature $T=(2\pi R_{5+1})^{-1}$, there are two critical values of $(\ell_0/(2\pi R_{5+1}))_{\rm{cr}} \approx 0.168$ 
and $0.175$. For $\ell_0/(2\pi R_{5+1})< 0.168$ there is always one connected solution, for $0.168 <\ell_0/(2\pi R_{5+1})< 0.175$ there are two, and for $\ell_0/(2\pi R_{5+1})>0.175$ 
there exists none.  Comparing these results to the ones obtained for the zero-temperature 
$\D 5$-$\D 7$-$\overline \D 7$ configuration, one observes that while at zero temperature \cite{ahk0608}, there are either an infinite number of solutions or none, at $T=(2\pi R_{5+1})^{-1}$ there are different numbers of solutions (one, two, or none) depending 
on what values $\ell_0$ take; see Figures 2 and 3. For $\D 5$-$\D
7$-$\overline \D 7$ configuration at temperatures $T<(2\pi R_{5+1})^{-1}$, the situation is the same as the zero temperature case. There exist connected solutions only for a specific value of $\ell_0=\pi R_{5+1}/3$. Indeed, there are an infinite number of such solutions; see Figure 3. 

For $\D 6$-$\D 6$-$\overline \D 6$ configuration, the situation is 
simpler. For any $\ell_0/\beta$ there always exists one and
only one connected solution.  As we will see in the next section all
these connected solutions are less energetically favorable compared to disjoint
solutions. The observation that for color $\D 6$-branes at finite
temperature there is always one connected solution, hence no "screening
length", is in accord with the fact that the near horizon geometry of
$\D 6$-branes cannot be decoupled from the gravitational modes of the 
bulk geometry \cite{imsy9802}.

\paragraph{$\bold{(1+1)}$-dimensional intersections} 
In this case, there are five allowed configurations (shown on the second and third rows
in Figure 2), namely $\D 2$-$\D 8$-$\overline \D 8$, $\D 3$-$\D 7$-$\overline 
\D 7$, $\D 4$-$\D 6$-$\overline \D 6$, $\D 5$-$\D 5$-$\overline \D 5$ and 
$\D 6$-$\D 4$-$\overline \D 4$. Configurations 
with $q\leq 4$ show similar behavior. For $q\leq 4$ there
always exists a critical $(\ell_0/\beta)_{\rm cr}$ beyond which
connected solutions cease to exist. The critical value obtained from
Figure 2 is  $(\ell_0/\beta)_{\rm cr}\approx 0.225, 0.223, 0.227$ for color 
$\D 2$, $\D 3$ and $\D 4$-branes, respectively. Below these
critical values there are two solutions, the short and long
solutions\footnote{The authors of \cite{gxz0605} studied the $\D 2$-$\D
8$-$\overline \D 8$ system at finite temperature but missed the
existence of the long connected solution.}.

For the $\D 5$-$\D 5$-$\overline \D 5$ system at $T=(2\pi R_{5+1})^{-1}$, below a critical value of $(\ell_0/(2\pi R_{5+1}))_{\rm cr}\approx 0.251$ there is always one
solution whereas above this value connected solutions do not exist.  Note the difference (depicted in Figure 3) of this case with the $\D 5$-$\D 7$-$\overline \D 7$ system: for the  $\D 5$-$\D 5$-$\overline \D 5$ configuration, there is no range of $\ell_0/(2\pi R_{5+1})$ for which there exist the short and long solutions. Like the $\D 5$-$\D 7$-$\overline \D 7$ configuration, for the $\D 5$-$\D 5$-$\overline \D 5$ system at $T<(2\pi R_{5+1})^{-1}$, there is an infinite number of connected solutions when $2\ell_0=\pi R_{5+1}$ and none for other $\ell_0$'s.

For the $\D 6$-$\D 4$-$\overline \D 4$ configuration, there is always
one connected solution for arbitrary values of $\ell_0/\beta$. Notice that the
behavior of the flavor branes in the geometry of color $\D 5$ and $\D 6$-branes with $(1+1)$-dimensional intersections is qualitatively the same as their behavior with $(3+1)$-dimensional intersections. 

\myfig{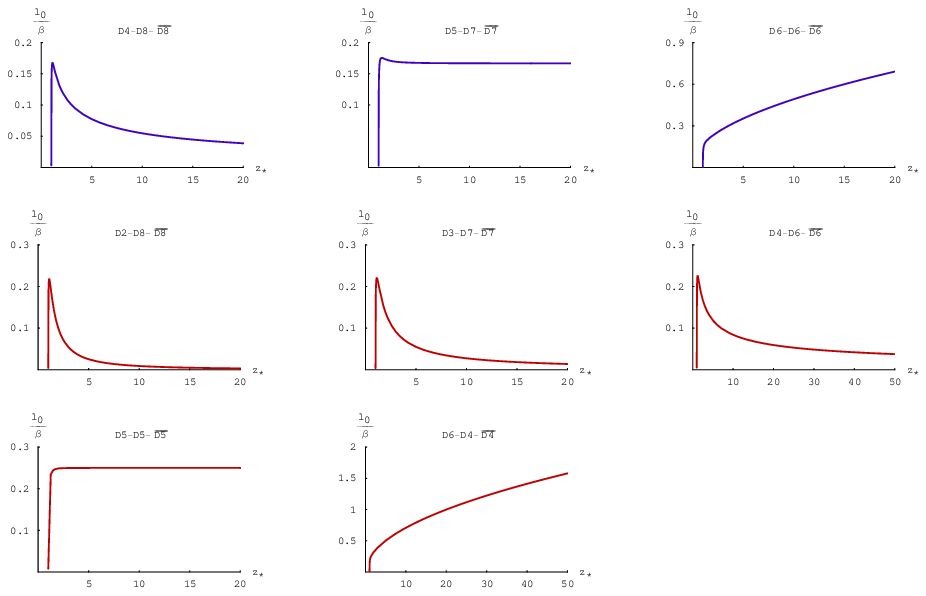}{15}{\footnotesize{Behavior of $\ell_0/\beta$ versus $z_*$ obtained 
numerically for generic values $z_*$ for various $q$'s and intersections 
of interest, i.e. $r=1$ and $r=3$. The blue plots show the behavior for $3+1$ 
intersections whereas the red plots show the behavior for $1+1$ intersections. These plots show the number of connected $\Dp$-branes (for all temperatures) placed in the background of black $\Dq$-branes. For $q=5$, the temperature $T$ is fixed: $T=(2\pi R_{5+1})^{-1}$. Except for $q=5, 6$, such solutions can potentially be realized as different phases of the holographic dual theories.}}

\myfig{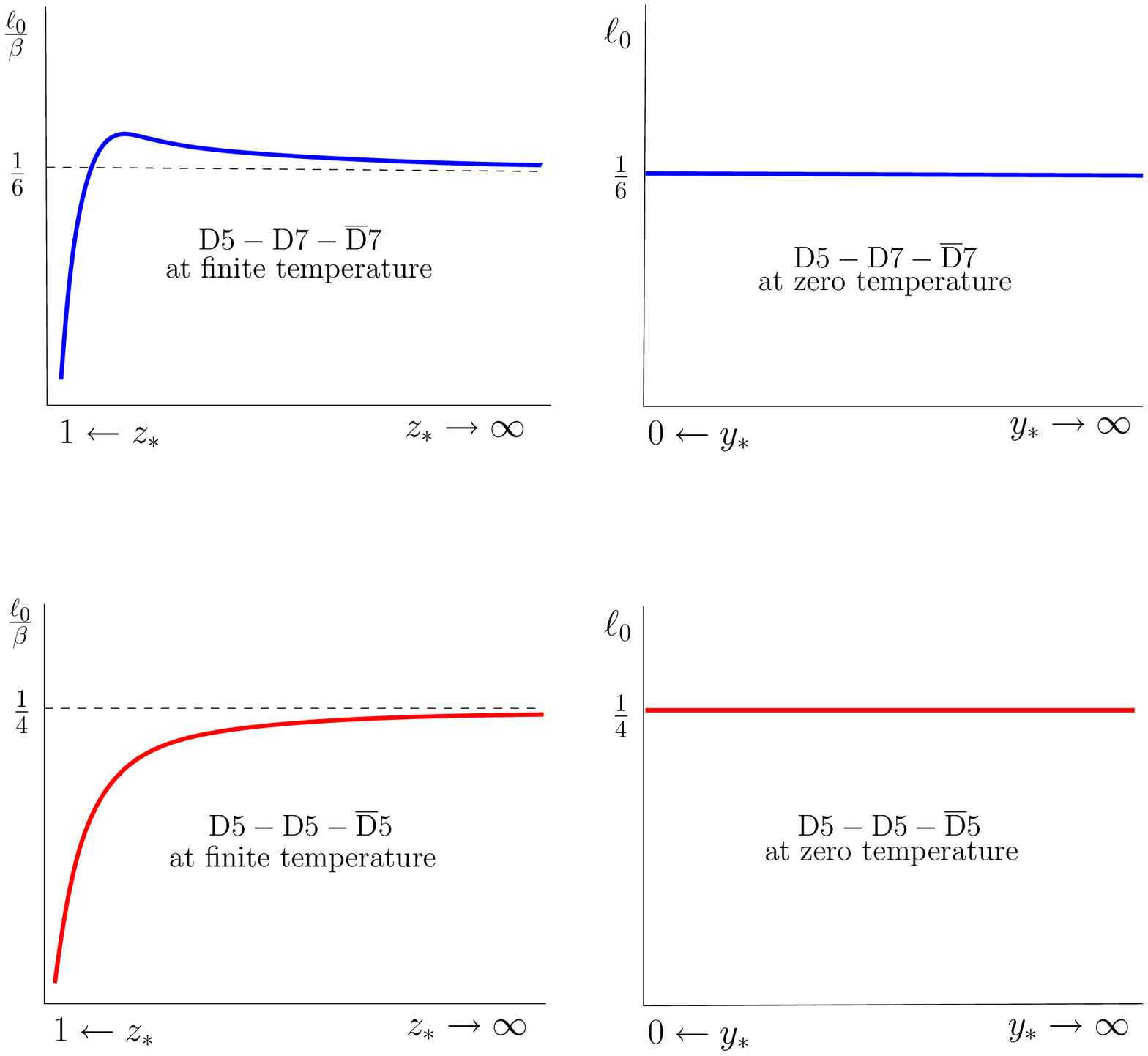}{10}{\footnotesize{The two plots on the left hand side show the number of connected solutions for the flavor $\D7$ and $\D5$-branes in the background of black $\D 5$-branes at $T=(2\pi R_{5+1})^{-1}$. The plots on the right hand side, on the other hand, show the number of connected solutions for the same flavor branes in the background of thermal (or zero temperature) $\D5$-branes. $z_*$'s represent the turning points of the connected flavor branes in the black $\D 5$-brane geometry while $y_*$'s are the turning points in the background of thermal $\D5$-branes.}}

\section{Energy of the configurations}

\myfig{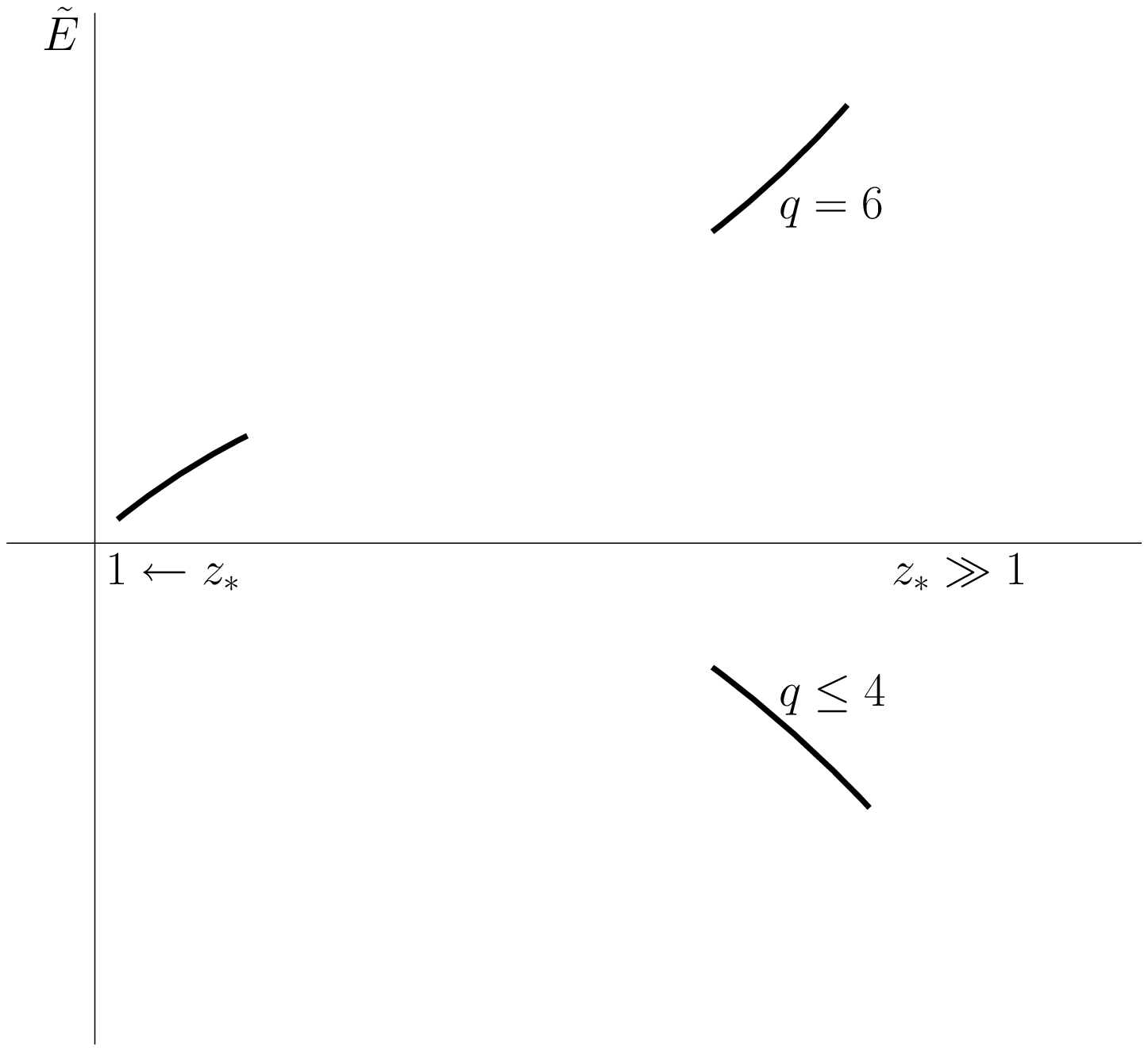}{7}{\footnotesize{Behavior of $\tilde E$ versus $z_*$ 
obtained numerically for generic values of $z_*$ for various $q$'s.}}

As we saw in the previous section there are typically more than one
solution for a given $\ell_0/\beta$. In fact, just to recap, for
$\ell_0/\beta$ less than a critical value there are generically three
branches of solutions for $q\leq 4$; one disjoint and two connected
solutions. There are also an infinite number of solutions for $q=5$ at $T<(2\pi R_{5+1})^{-1}$ as long as $\ell_0=2\pi R_{5+1}/(r+3)$. There
are two branches of solutions for $q=6$ for any $\ell_0/\beta$; one
disjoint and one connected solution. Since there are various
configurations for a particular value of $\ell_0/\beta$, one needs to
compare their on-shell actions to determine which configuration is more
energetically favorable. In this section we analyze the energy of these
configurations by a combination of analytical and numerical techniques. The energy of these configurations by themselves is infinite. We regulate the energies of connected
configurations by subtracting from them the energy of disjoint
configurations
\bea\label{e}
\tilde E= {\rm{lim}}_{\Lambda \to \infty}\left\{\int_{z_*}^\Lambda
dz~z^{\frac{\sigma}{2}}
\Big(1-\frac{z_*^\gamma-z_*^\sigma}{z^\gamma-z^\sigma}\Big)^{-\frac{1}{2}}
- \int_{1}^\Lambda dz~z^{\frac{\sigma}{2}}\right\},
\eea
where $\Lambda$ is a cutoff and the difference in energy $E$ is related to $\tilde E$ as
\bea 
E= -\frac{\delta \beta^2}{\pi}
C(q,r)~u_T^{\frac{\gamma}{2}} \int d^rx ~{\tilde E}.
\eea

\subsection{Analytical analysis}
The integral in (\ref{e}) is complicated and as far as we know cannot be
integrated analytically for generic values of $z_*$. However, like the
integral in (\ref{Lbetau0eq}), there are two regions of $z_*\to 1$ and
$z_*\gg 1$ for which we can integrate ${\tilde E}(z_*)$ analytically. In
the $z_* \to 1$ limit, it can be shown that $\tilde E$ is positive for
all $q$ and $r$.  Indeed, let's rewrite (\ref{e}) as follows
\bea\label{e-smallzstar-1}
\tilde E_\Lambda &=& \int_{z_*}^\Lambda
dz~z^{\frac{\sigma}{2}}\Big[\Big(1-\frac{z_*^\gamma-z_*^\sigma}
{z^\gamma-z^\sigma}\Big)^{-\frac{1}{2}}-1\Big]- \int_{1}^{z_*}
dz~z^{\frac{\sigma}{2}}\nonumber\\
&=& \Big{(}\int_{z_*}^{z'_*}
dz~z^{\frac{\sigma}{2}}\Big[\Big(1-\frac{z_*^\gamma-z_*^\sigma}
{z^\gamma-z^\sigma}\Big)^{-\frac{1}{2}}-1\Big]- \int_{1}^{z_*}
dz~z^{\frac{\sigma}{2}}\Big{)} \nonumber\\
& &\hspace{100pt}+ \int_{z'_*}^\Lambda
dz~z^{\frac{\sigma}{2}}\Big[\Big(1-\frac{z_*^\gamma-z_*^\sigma}
{z^\gamma-z^\sigma}\Big)^{-\frac{1}{2}}-1\Big].
\eea
The last integral in (\ref{e-smallzstar-1}) is positive for any $z'_*$.
If $z_* = 1 + \epsilon$, we can choose $z'_* = 1 + 3\epsilon$ and the
difference in the bracket is estimated to be
\bea\label{e-smallzstar-2} 
\epsilon(\sqrt{6}-2) +
\epsilon\Big(\log(\sqrt{2}+\sqrt{3})-1\Big) + o(\epsilon^2), 
\eea
which is also positive.

Approximating (\ref{e}) in the $z_* \gg 1$ region yields
\bea\label{e-largezstar-1}
 {\tilde E}_\Lambda &\sim& \int_{z_*}^\Lambda dz~z^{\delta+\gamma}
\Big(z^\gamma-z_{*}^\gamma\Big)^{-\frac{1}{2}} - \int_{1}^\Lambda dz~z^{\frac{\sigma}{2}},\nonumber\\
&=& \left\{\int_{z_*}^\Lambda dz~z^{\delta+\gamma}
\Big(z^\gamma-z_{*}^\gamma\Big)^{-\frac{1}{2}} - \int_{0}^\Lambda
dz~z^{\frac{\sigma}{2}}\right\} + \frac{2}{\sigma+2}.
\eea

\myfig{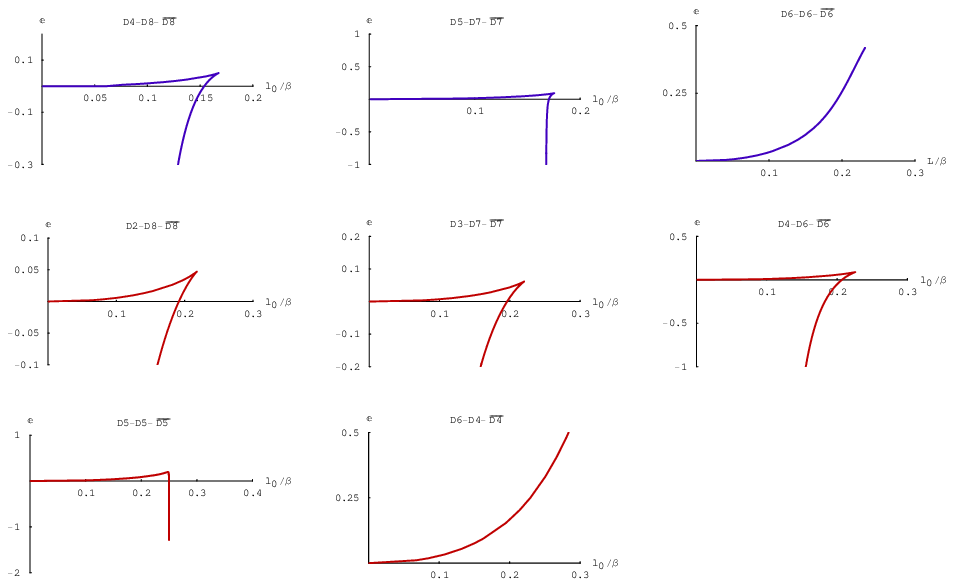}{15}{\footnotesize{Behavior of $\tilde E$ versus $\ell_0/\beta$ obtained numerically for
various $q$'s and intersections of interest, i.e. $r=1$ and $r=3$. The blue plots show the behavior for $3+1$ intersections whereas the red plots show the behavior for $1+1$ intersections. The plots for the color $\D5$-branes are for a fixed (inverse) temperature of $\beta=2\pi R_{5+1}$.}}

Note that in the large $z_*$ limit, $z_* \simeq z_0$ where $z_0=u_0/u_T$. As a result, from
(\ref{e-largezstar-1}) we see that the energy of the connected
configurations approaches their zero temperature value obtained in
\cite{ahk0608}. This is expected since in this regime the connected
flavor branes are very far away from the horizon hence receive little
effect from it. The disjoint configuration, on the other hand, always
keeps in touch with the horizon and gets a finite temperature contribution
$-\frac{2}{\sigma+2}$. The term in the curly bracket of
(\ref{e-largezstar-1}) is already computed in \cite{ahk0608} in terms of
Beta functions, giving an energy difference of
\bea\label{e-largezstar-2} 
{\tilde E} &=& {\rm{lim}}_{\Lambda \to \infty} {\tilde E}_\Lambda\nonumber \\
&=& \frac{1}{\gamma} z_*^{\frac{\gamma}{2} + \delta + 1}
B\Big[-\frac{1}{2}-\frac{\delta + 1}{\gamma},\frac{1}{2}\Big]+
\frac{2}{\sigma+2}. 
\eea 
Since $z_* \gg 1$, the last term is irrelevant in determining the sign
of ${\tilde E}$. We see immediately that ${\tilde E} > 0$ for $q\geq 6$
and ${\tilde E} < 0$ for $q\leq 4$. For $q=5$, the Beta function
vanishes, and a more careful investigation must be made to
determine the sign of ${\tilde E}$. It turns out that in going from
(\ref{e}) to (\ref{e-largezstar-1}), we have over-estimated the energy
for the joined brane configurations: there is a correction with leading
behavior $\sim -z_*^{(r-1)/2}$ for large $z_*$. Thus, for $q=5$ we
have ${\tilde E} < 0$ for $r > 1$, while the analytic analysis is not
reliable for $r=1$. The analytic results for the two aforementioned
limits of $z_*$ are summarized in Figure 4.

\subsection{Numerical results and phase transitions} 
For generic value of $z_*$ the integral in (\ref{e}) can be
numerically integrated and one can plot ${\tilde E}$ versus $z_*$.
Instead we will numerically eliminate $z_*$ between (\ref{Lbetau0eq}) 
and (\ref{e}) and plot ${\tilde E}$ versus $\ell_0/\beta$. The reason
for doing so is that one can easily observe the transitions, say
from a connected configuration to a disjoint configuration, in terms of
both $\ell_0$ and $\beta$. The numerical results are shown in Figure 5.
In the following we will analyze these graphs and determine the phases
of the vacuum as one varies $\ell_0/\beta$.

\paragraph{$\bold{(3+1)}$-dimensional intersections} 
As mentioned in the
previous section, for $3+1$-dimensional intersections we have three
configurations. Consider first
the $\D 4$-$\D 8$-$\overline \D 8$ system.  For $\ell_0/\beta<
(\ell_0/\beta)^{\rm e}_{\rm cr}\approx 0.154$ the short solution is more
energetically favorable compared to both the disjoint and the long
solutions. This corresponds to the chiral symmetry being broken. (Here
$(\ell_0/\beta)^{\rm e}_{\rm cr}$ is a critical value read off the
energy diagrams in Figure 5. It is different from $(\ell_0/\beta)_{\rm
cr}$ obtained from the plots of Figure 2.) Above this value (up to
$\ell_0/\beta \approx 0.17$) both short and long curved solutions have
more energy compared to the disjoint solution, indicating that the 
disjoint solution is the vacuum, hence chiral symmetry is restored and 
the phase transition is of first order. Thus the chiral symmetry
breaking-restoration phase transition occurs at $(\ell_0/\beta)^{\rm
e}_{\rm cr}\approx 0.154$.

For the $\D 5$-$\D 7$-$\overline \D 7$ system at $T=(2\pi R_{5+1})^{-1}$ the behavior is 
surprising and to some extent more involved. On the $\D 5$-$\D 7$-$\overline \D 7$ plot in Figure 5 there are two special points with $(\ell_0)^{\rm e}_{\rm cr}\approx 0.168$
and $0.17$ (in units of $2\pi R_{5+1}$). For $\ell_0/(2\pi R_{5+1})<0.168 $ the disjoint solution is more energetically favorable, hence chiral symmetry in the dual theory is intact given that such a solution can represent a valid phase of the dual theory. For $0.168 <
\ell_0/(2\pi R_{5+1})<0.17$ there are three kinds of solutions, disjoint, short
and long. It turns out that the short solution has less energy than the
other two indicating that chiral symmetry is broken. For $\ell_0/(2\pi R_{5+1})>0.17$ the disjoint solution becomes more energetically
favorable, hence potentially chiral symmetry gets restored. For temperatures less than $(2\pi R_{5+1})^{-1}$, the situation is the same as the zero temperature case: there are an infinite number of connected solutions when $3\ell_0=\pi R_{5+1}$, each equally energetically favored, and each of them more favored over the disjoint solution.  The existence of such solutions may be rooted in the fact that the low energy theory on the color $\D 5$-branes is a non-local field theory, a little string theory. Due to the fact that in the case of background $\D5$-brane geometry the dual field theory degrees of freedom cannot be totally decoupled from non-field theoretic degrees of freedom, it is not clear to us whether such solutions can represent a chirally-broken phase of the dual gauge theory despite the fact that, geometrically, they are smoothly connected. 

For the $\D6$-$\D 6$-$\overline \D 6$ system, the disjoint solution is
always favorable, although it is not clear whether one can  give a
holographic interpretation that the ``dual" field theory is in a chirally-symmetric phase. This is because there exists no decoupling limit suitable for holography in the case of background $\D6$-branes.

\paragraph{$\bold{(1+1)}$-dimensional intersections} 
For
(1+1)-dimensional intersections, there are five allowed configurations.
Among these the models based on background $\Dq$-branes with $q\leq 4$ exhibit
similar behaviors to their counterparts with (3+1)-dimensional intersection. That is to say for sufficiently low temperatures (compared to
$1/\ell_0$) chiral symmetry is broken while above a critical temperature
it gets restored. The critical values at which this (first order) phase 
transition occurs are $(\ell_0/\beta)^{\rm e}_{\rm cr}\approx 0.191$, 
$0.196$ and $0.206$ for color $\D 2$, $\D 3$ and $\D 4$-branes, 
respectively.

The model with color $\D5$-branes again shows some surprises. Because of the mixing between the field theoretic and non-field theoretic degrees of freedom in holography involving background $\D5$-branes, we have no evidence that different behaviors of the flavor branes represent, via holographic point of view, either chirally-symmetric or chirally-broken phases of the dual field theory. Nevertheless, one finds the following results. At  $T=(2\pi R_{5+1})^{-1}$ and for small enough $\ell_0$ ($\ell_0< (\ell_0)^{\rm e}_{\rm
cr}\approx 0.2498 \times 2\pi R_{5+1}$) the disjoint solution is preferred. Increasing 
$\ell_0$ up to $\ell_0/(2\pi R_{5+1})<0.251$ will result in a phase where the connected solution is favorable. This phase appears in our plot because we put 
a cutoff of $\Lambda = 5$ to regulate the energy integral.
Increasing the cutoff will decrease the range of
$\ell_0$ for which this phase exists. It is plausible that in the 
$\Lambda \to \infty$ limit this phase disappears although our numerics does not allow us to check this explicitly. Sticking for now with the cutoff we chose, if one increases
$\ell_0$ further, there will be another phase transition to a phase where 
the disjoint solution becomes favored.  Due to space limitations, the
resolution of the $\D 5$-$\D 5$-$\overline \D 5$ plot in Figure 5 does
not allow one to see all these phases. For temperatures less than 
$(2\pi R_{5+1})^{-1}$, like the  zero temperature case, there are an 
infinite number of connected solutions for $\ell_0=\pi R_{5+1}/2$, each 
equally energetically favored, and none for other $\ell_0$'s. Each 
of these connected solutions is more favored over the disjoint solution. For 
the $\D 6$-$\D 4$-$\overline \D 4$ system, there is no phase
transition and it is always the disjoint solution which is energetically
favorable.

\section{Transverse intersections at finite temperature with compact $x^q$}

An interesting property of the models we are studying here is that when
$x^q$ is compact the scale of chiral symmetry breaking is generically
different from the scale of confinement which results in additional
phases.  For example, for the Sakai-Sugimoto model at finite
temperature, it was shown \cite{asy0604} that there exists an
intermediate phase where the system is deconfined while chiral symmetry
is broken.

At finite temperature and $x^q$ direction being compact (with a radius 
of $R_c$), there are three geometries where the topology of the 
$t-x^q$ submanifold is ${\rm S}^1\times {\rm S}^1$. One is a geometry 
which has the metric 
\bea\label{compactmetric1}
ds^2=\left(\frac{u}{R_{q+1}}\right)^{\frac{7-q}{2}}\Big(dt^2 + d{\vec x} ^2 +g(u) ({dx^q})^2
\Big) + \left(\frac{u}{R_{q+1}}\right)^{-\frac{7-q}{2}}
\Big(\frac {du^2}{g(u)}+ u^2 d{\Omega_{8-q}}^2\Big),
\eea
with
\bea\label{warpfactor1}
g(u)= 1-\Big(\frac{u_{\rm KK}}{u}\Big)^{7-q}.
\eea
We will call this geometry the thermal geometry. Although the Euclidean time period $\beta$ is arbitrary in this geometry,
the $x^q$-circle cannot have arbitrary periodicity. In order for this
geometry to be smooth at $u=u_{\rm KK}$ in the $x^q-u$ submanifold, one
has to have
\bea\label{xqperiod}
\Delta x^q=\beta_c=\frac{4\pi}{7-q}\left(\frac{R_{q+1}}{u_{\rm KK}}\right)^{\frac{7-q}{2}}u_{\rm KK},
\eea
where we have defined $\beta_c=2\pi R_c$. Although we do not specify the $q$-dependence of $\beta_c$ and $R_c$, one should keep in mind that they depend on $u_{\rm KK}$ and $R_{q+1}$ differently through (\ref{xqperiod}) depending on what value for $q$ is given. There is another geometry whose metric takes the form
\bea\label{compactmetric3}
ds^2=\left(\frac{u}{R_{q+1}}\right)^{\frac{7-q}{2}}\Big(dt^2 + d{\vec x} ^2 +(dx^q)^2
\Big) + \left(\frac{u}{R_{q+1}}\right)^{-\frac{7-q}{2}}
\Big(du^2+ u^2 d{\Omega_{8-q}}^2\Big).
\eea
There is also the black brane geometry which is basically the same as (\ref{Dq metric}) but
with $x^q$ compact, and has the (Euclidean) metric
\bea\label{compactmetric2}
ds^2=\left(\frac{u}{R_{q+1}}\right)^{\frac{7-q}{2}}\Big(f(u) dt^2 + d{\vec x} ^2 + ({dx^q})^2
\Big) + \left(\frac{u}{R_{q+1}}\right)^{-\frac{7-q}{2}}
\Big(\frac {du^2}{f(u)}+ u^2 d{\Omega_{8-q}}^2\Big),
\eea
where
\bea\label{warpfactor2}
f(u)= 1-\Big(\frac{u_T}{u}\Big)^{7-q}.
\eea
In this geometry the $x^q$-circle has arbitrary periodicity whereas $\beta$ is fixed by 
\bea\label{timeperiod}
\beta=\frac{4\pi}{7-q}\Big(\frac{R_{q+1}}{u_T}\Big)^{\frac{7-q}{2}}u_T.
\eea
For all three geometries, the dilaton $\phi$, and the $q$-form RR-flux $F_q$
are given in (\ref{dilatonq-form}); see the appendix for more details. Also, $R_{q+1}$ is given in
(\ref{radius}).

The three geometries whose line elements are given in (\ref{compactmetric1}), (\ref{compactmetric3}) and (\ref{compactmetric2}) are saddle points of either type IIA or type IIB Euclidean path integral.
For a given temperature, one needs to compare their (regularized) free
energies to determine which solution dominates the path integral. We have calculated the free energies of these solutions in the appendix. For $q\leq 4$, both the thermal and the black brane geometries have less free energy compared to the geometry in (\ref{compactmetric3}), and this result is independent of temperature. So, to determine the lowest energy saddle point we need to compare  the free energies of (\ref{compactmetric1}) and (\ref{compactmetric2}). The difference in their free energies is given by
\bea
S_{\rm thermal}-S_{\rm black~brane}=\frac{9-q}{g_s^2} V_9 \left(\frac{4\pi}{7-q}R_{q+1}^{\frac{1}{2}(7-q)} \right)^{2\frac{7-q}{5-q}}\left(\beta^{2\frac{q-7}{5-q}}-\beta_c^{2\frac{q-7}{5-q}}\right),
\eea
where $V_9=\beta \beta_c {\rm Vol}({\rm S}^{8-q}){\rm Vol}(\mathbb{R}^{q-1})$ where we set $l_s=1$. Thus the thermal geometry (\ref{compactmetric1})
dominates when $\beta>2\pi R_c$ whereas for $\beta<2\pi R_c$ it is the black brane geometry
(\ref{compactmetric2}) which gives the dominant contribution to the
Euclidean path integral. This phase transition which happens at $\beta=\beta_c$ is the holographic dual of confinement-deconfinement phase transition in the corresponding dual gauge theories \cite{w9803}.

For $q=6$, again, both the thermal and black brane geometries have less free energy 
compared to the geometry in (\ref{compactmetric3}). The difference in free energies for the geometries in  (\ref{compactmetric1}) and (\ref{compactmetric2}) is given by 
\bea\label{D6trans1}
S_{\rm thermal~\D6}-S_{\rm black~\D6-brane}= \frac{3V_9}{16\pi^2 g_s^2R_{6+1}} (\beta^2-\beta_c^2), 
\eea
where $V_9=\beta \beta_c {\rm Vol}({\rm S}^{2}){\rm Vol}(\mathbb{R}^{5})$. So, (\ref{D6trans1}) indicates a phase transition at $\beta=2\pi R_c$. Unlike the $q\leq 4$ cases,  for $q=6$ the thermal geometry (\ref{compactmetric1})
dominates for $\beta<2\pi R_c$ whereas for $\beta>2\pi R_c$ the black brane geometry
(\ref{compactmetric2})  dominates. 

For $q=5$, one needs to consider more possiblities. For $\beta=2\pi R_{5+1}$, there are two cases: either $\beta_c=2\pi R_{5+1}$ or $\beta_c\neq2\pi R_{5+1}$. For $\beta_c=2\pi R_{5+1}$, both (\ref{compactmetric1}) and (\ref{compactmetric2}) are more favored over the geometry whose metric is given in (\ref{compactmetric3}). The difference in free energies of  (\ref{compactmetric1}) and (\ref{compactmetric2}) is 
\bea
S_{\rm thermal~\D5}-S_{\rm black~\D5-brane}= \frac{4}{g_s^2} V_9 (u_T^2-u_{\rm KK}^2), \qquad {\rm at} \qquad \beta_c=2\pi R_{5+1}.
\eea
where $V_9=\beta \beta_c {\rm Vol}({\rm S}^{3}){\rm Vol}(\mathbb{R}^{4})$. For $\beta_c\neq2\pi R_{5+1}$, on the other hand, the two saddle points with the same asymptotics are the black brane geometry and the geometry in (\ref{compactmetric3}). In this case, the black brane geometry is the dominant one
\bea
\Delta S=  \frac{4}{g_s^2} V_9 u_T^2 >0.
\eea
There are also two possibilities when $\beta\neq2\pi R_{5+1}$. If $\beta_c\neq2\pi R_{5+1}$, the only saddle point consistent with the asymptotics is (\ref{compactmetric3}) which determines the vacuum. If, on the other hand, $\beta_c=2\pi R_{5+1}$, there are two geometries with the same asymptotics, thermal and the one given in (\ref{compactmetric3}). The thermal geometry is dominant because 
\bea
\Delta S= -\frac{4}{g_s^2} V_9 u_{\rm KK}^2 <0.
\eea

With $x^q$ being compact, the flavor branes are now sitting at two
points separated by a distance $\ell_0$ on the $x^q$-circle. Note 
that there is no reason for the flavors branes to be
located at the antipodal points on the circle.  As before, we consider
the transverse intersections of the flavor and the color branes,
and choose the same embeddings and boundary conditions for the flavor 
branes as we did in (\ref{embedding}) and (\ref{Dbranebc1}) except 
that now $\ell_0 \leq \pi R_c$. In what follows, we will focus on $q\leq4$ 
cases and consider both low and high temperature phases of the 
background. In particular, we would like to know whether there always 
exists a range of temperature above the deconfinement temperature 
where chiral symmetry is broken.

\subsection{Behavior at high and low temperatures}
For temperatures above the deconfinement temperature $\beta_c$ the profile of the 
flavor branes takes essentially the same form as it did when $x^q$ was non-compact, 
hence, indicating the existence of short and long (smoothly) connected solutions.   
Note that when $x^q$ is compact, for a fixed $\ell_0$, there is now a 
lower bound on $\ell_0 /\beta$ set by the deconfinement temperature. 
For completeness, we have plotted $\ell_0 /\beta$ 
versus $z_*$ ($z_*$ being the radial position at which the brane and 
anti-brane smoothly join) in Figure 6. The lower dotted line in each plot 
represents the deconfinement 
temperature. As an example, we chose it to be at $\beta_c=10 \ell_0$. The 
upper dotted line shows chiral symmetry breaking-restoration phase 
transition which comes from comparing the energies of connected and disjoint 
solutions. One can show, using energy 
considerations, that below the upper dotted line chiral symmetry is broken  
in a deconfined phase via short connected solution while it is restored 
above the line (where we have deconfinement with chiral symmetry restoration). 

\myfig{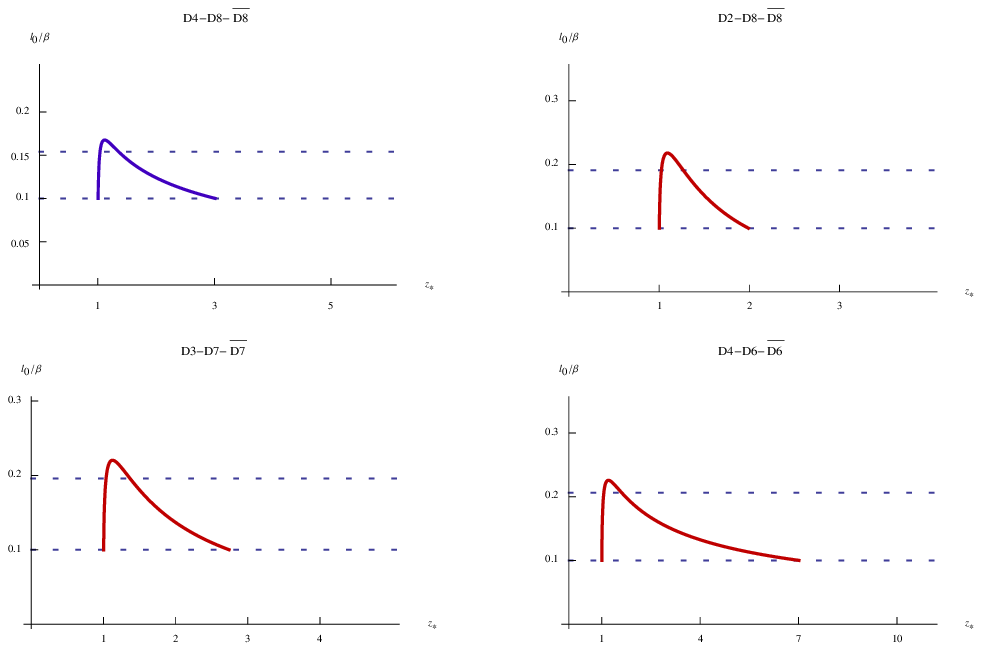}{16}{\footnotesize{Behavior of $\ell_0 /\beta$ versus $z_*$ 
above the deconfinement temperature for $q=2,3,4$. The lower dotted 
line represents a transition to the confined phase whereas the upper dotted 
line represents chiral symmetry breaking-restoration phase transition.}}

In the low temperature regime where the system is in a confined phase, 
the DBI action for the flavor branes in the thermal background 
(\ref{compactmetric1}) now reads (with gauge fields set equal to zero)
\bea\label{DBIcompactmetric1}
S_{\rm{DBI}}=\beta\,C(q,r)\int d^rx~dx^q ~ u^{\frac{\gamma}{2}} 
\Big[g(u)+\frac{1}{g(u)}\Big(\frac{u}{R_{q+1}}\Big)^{2\delta}{u^{'}}^2\Big]^{\frac{1}{2}},
\eea
where $C(q,r)$, $\gamma$ and $\delta$ have all been defined in 
(\ref{parameters}). The equation of motion for the profile is now
\bea\label{compact1eom}
{u^{\frac{\gamma}{2}} g(u)}\Big[g(u)+\frac{1}{g(u)}\Big(\frac{u}{R_{q+1}}\Big)^{2\delta}{u^{'}}^2\Big]^{-\frac{1}{2}}=w_0^{\frac{\gamma}{2}},
\eea
with $w_0$ parameterizing the solutions. There exists a solution 
with $w_0=0$ representing disjoint $\Dp$ and $\bDp$-branes 
descending down to $u=u_{\rm KK}$. For $w_0\neq 0$, solving 
(\ref{compact1eom}) for $u^{'}$ yields 
\bea\label{compact1uprime}
{u^{'}}^2=\frac{1}{{w_0}^\gamma} \Big(\frac{u}{R_{q+1}}\Big)^{-2\delta} g(u)^2~\Big(u^\gamma g(u) -{w_0}^\gamma\Big).
\eea 
Denoting the possible turning point(s) by  $w_{*}$, analysis 
of (\ref{compact1uprime}) shows that $u\geq w_{*}>u_{\rm KK}$, 
with $w_{*}$ satisfying 
\bea\label{compact1turningpoint}
w_{*}^\gamma-w_{*}^\sigma u_{\rm KK}^{-2\delta}-w_0^\gamma =0,
\eea
where $\sigma$ has been defined as before. Integrating (\ref{compact1uprime}) 
gives
\bea\label{xvsycompact}
x^q(y)=-\frac{\delta}{2\pi} \beta_c (y_{*}^\gamma-y_{*}^\sigma)^{\frac{1}{2}}\int_{y_{*}}^y \Big({\tilde y}^{-2\delta}-1\Big)^{-1}
\Big({\tilde y}^\gamma-{\tilde y}^\sigma -(y_{*}^\gamma-y_{*}^\sigma)\Big)^{-\frac{1}{2}} d{\tilde y},
\eea
where we have defined $y=(u/u_{\rm KK}) \in (1,\infty)$, and 
$y_*=w_*/u_{\rm KK}$. Using (\ref{xvsycompact}) we can relate $y_*$ to 
$\ell_0$
\bea\label{Lvsweq}
\frac{\ell_0}{\beta_c}=-\frac{\delta}{\pi}~(y_{*}^\gamma-y_{*}^\sigma)^{\frac{1}{2}}\int_{y_{*}}^{\infty} \Big({y}^{-2\delta}-1\Big)^{-1}
\Big({y}^\gamma-{y}^\sigma -(y_{*}^\gamma-y_{*}^\sigma)\Big)^{-\frac{1}{2}} dy.
\eea
The analysis of $\ell_0/\beta_c$
as a function of $y_*$ determines the number of solutions. 
$\ell_0/\beta_c$ versus $y_*$ 
has been numerically plotted in Figure 7 for intersections of 
interest and for $q=2,3,4$. As it is seen from Figure 7, 
there is always one smoothly connected solution (as well as a disjoint 
solution). The red and blue plots represent smoothly connected solutions 
for (1+1)-dimensional and (3+1)-dimensional intersections, 
respectively. 
\myfig{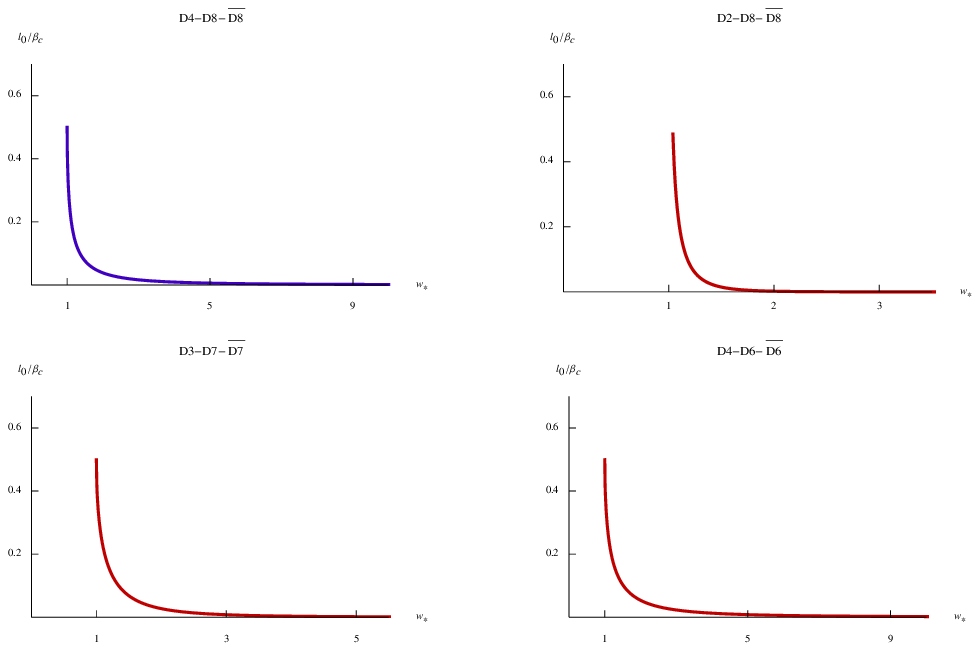}{15}{\footnotesize{Behavior of $\ell_0 /\beta_c$ versus $w_*$ for $q=2,3,4$.}}

For the energy of the connected solutions, one obtains
\bea 
E= -\frac{\delta \beta }{\pi}\beta_c
C(q,r)~u_{\rm KK}^{\frac{\gamma}{2}} \int dx^r ~{\tilde E},
\eea
where 
\bea\label{compact1e}
\tilde E= {\rm{lim}}_{\Lambda \to \infty}\left\{\int_{y_*}^\Lambda
dy~y^{\frac{\sigma}{2}} \Big(1-y^{2\delta}\Big)^{-\frac{1}{2}}
\Big(1-\frac{y_*^\gamma-y_*^\sigma}{y^\gamma-y^\sigma}\Big)^{-\frac{1}{2}}
- \int_{1}^\Lambda dy~\Big(1-y^{2\delta}\Big)^{-\frac{1}{2}}y^{\frac{\sigma}{2}}\right\},
\eea
and $\Lambda$ is a cutoff. Like the previous sections, one can numerically 
eliminate $y_*$ between (\ref{Lvsweq}) and (\ref{compact1e}) and plot 
${\tilde E}$ versus $\ell_0/\beta_c$. Although we have not shown the plots 
here, one can check (numerically) that the smoothly connected solutions are always more 
energetically favorable compared to the disjoint solutions.

The plots in Figure 
7 not only show the existence of a unique smoothly coonected solution (for 
the confined phase) but also indicate a big difference for the 
behavior of the flavor branes below and above the confinement-deconfinement 
phase transition. For example consider the plot for $\D 3$-
$\D 7$-$\overline \D 7$ system. Each point on the plot represents 
a unique curved solution for which increasing the temperature (up 
to the deconfinement temperature $\beta_c$) will have no effect on the shape 
of the U-shaped flavor branes. That is to say that changing the 
temperature will not cause the flavor $\D 7$-$\overline \D 7$-branes 
to join either closer to $u_{\rm KK}$ or farther away from it. Once 
$\ell_0$ and $\beta_c$ are specified, the shape of the brane stays 
the same independent of the variation of the temperature up to the 
deconfinement temperature. The behavior just mentioned is significantly 
different from the behavior of the flavor branes at high temperatures (where 
they are in black brane backgrounds). For a fixed $\ell_0$, each 
point on the plot of $\D 3$-$\D 7$-$\overline \D 7$ system in Figure 7 
represents one (or two) smoothly connected solution(s) for only a specific 
temperature. Varying the temperature will now change the shape of 
the connected flavor branes and force them to go either closer to 
the horizon or stay farther away from it.  

There is another difference which is worthy of mentioning here. At 
the temperature for which the confinement-deconfinement phase transition 
occurs, namely at $\beta=\beta_c$, $u_{\rm KK}=u_T$. For a 
fixed $\ell_0$, flavor branes at temperatures above $\beta_c$ join 
at a radial point closer to $u_{\rm KK}=u_T$ than the same flavor 
branes placed at temperatures below $\beta_c$.

\section{Discussion}
In this paper, we analyzed some aspects of transversely-intersecting 
$\Dq$-$\Dp$-$\bDp$-branes at finite temperature. In particular, we mapped out 
different vacuum configurations which can holographically be identified 
with chiral symmetry breaking (or restoration) phase of their holographic 
dual theories. Although we showed that generically the long connected solutions 
are less energetically favorable compared to the short connected solutions 
we did not discuss their stability against small perturbations. Presumably 
a stability analysis along the lines of \cite{fgmp0704} can be done to 
show that the long connected solution is unstable against small perturbations. 
The analysis presented here can be generalized in various directions. For 
example, one can add a chemical potential to the setup and look for new phases 
as was done in some specific models in \cite{ht0608, bll0708, dgks0708, rsvw0708}, or consider 
the sysytem at background electric and magnetic fields \cite{bll0802, jk0803, 
ksz0803} and study the conductivity of the system or the effect of the magnetic 
field on the chiral symmetry-restoration temperature. We hope to come back 
to these interesting issues in future. 

Also, our analysis was entirely based on the DBI action for
the flavor $\Dp$-${\bDp}$-branes. In transversely-intersecting D-branes,
working with just the DBI (plus the Chern-Simons part of the) action 
misses, from holographic perspectives, an important part of the physics, 
namely the vev of the fermion bilinear as 
an order parameter for chiral symmetry breaking. To determine the fermion 
bilinear from the holographic point of view there should be a mode 
propagating in the bulk geometry such that asymptotically its normalizable 
mode can be identified with the vev of the fermion bilinear. The DBI plus 
the Chern-Simons action cannot give rise to the mass of the localized 
chiral fermions either. Note that for transverse intersections 
one cannot write an explicit mass term for the fermions of the intersections 
because there is no transverse space common to both the color and the flavor 
branes. So it is not possible to stretch an open string between the 
color and flavor branes in the transverse directions. Therefore, the 
fermion mass must be generated dynamically. It has been argued in \cite{ahjk0604} 
that including the dynamics of an open string stretched between the flavor 
branes into the analysis will address the question of how one can
compute fermion mass and bilinear vev in these holographic models. More
concretely, the scalar mode of this open string which transforms as
bifundamental of $U(N_{\f})\times U(N_{\f})$ has the right quantum
numbers to be potentially holographically dual to the fermion mass and
condensation.

Recently, the authors of
\cite{ckp0702, bss0708, dn0708} shed light on this issue by starting
with the so-called tachyon-DBI action \cite{g0411} claimed to 
correctly incorporate the role of the open string scalar mode, the tachyon, 
in a system of separated $\Dp$-${\bDp}$-branes. In fact, it was shown in 
\cite{bss0708, dn0708} that for the Sakai-Sugimoto model at zero temperature,  
the open string tachyon will asymptotically have a normalizable as well 
as a non-normalizable mode. They identified the normalizable mode with the 
vev of the fermion bilinear (order parameter for chiral symmetry breaking) and the 
non-normalizable mode with the fermion mass. It is not hard to generalize 
the calculations of \cite{bss0708, dn0708} to include all transversely-intersecting 
$\Dq$-$\Dp$-${\bDp}$ systems at zero temperature where one finds 
that there always exist both normalizable and non-normalizable 
modes for the asymptotic behavior of the tachyon, and in the bulk of the 
geometry the tachyon condenses roughly at the same radial point where the 
flavor branes smoothly join \cite{eln}. The calculation of the fermion mass 
and condensate for transversely-intersecting $\Dq$-$\Dp$-${\bDp}$ systems 
at finite temperature requires not only considering the tachyon-DBI action of \cite{g0411} 
in the black brane background of (\ref{Dq metric}) but also calculating 
the tachyon potential as a function of the temperature. In the case of a 
coincident brane and anti-brane in flat background, a partial result for 
such a calculation was given in \cite{dgk0109} (see also \cite{h0212}). 
Some attempts in generalizing the results of \cite{dgk0109, h0212} for 
separated brane-anti-branes (at least in the case of separated 
$\D 8$-$\overline \D 8$ in flat space) has recently started in 
\cite{ct0802}. For our purpose of extracting information about fermion 
mass and condensate, knowing the dependence of the tachyon potential for 
separated flavor $\Dp$-${\bDp}$-branes seems crucial. Ignoring the temperature 
dependence on tachyon potential at zeroth order yields unsatisfactory results: 
It gives rise to the same results as one would have obtained in the 
zero temperature case \cite{eln}. We know that this is not the right 
behavior because at zero temperature there is no chiral phase transition.

\paragraph{Note added in version two} There are now two more interesting (and closely related) proposals on how to compute the fermion mass and condensates in transversely-intersecting D-branes. One proposal is based on an open wilson line operator whose vev gives the condensate \cite{ak0803} (see also \cite{mms0807, aelv0811} for generalizations) and the other \cite{hhly0803} is based on a D6-brane, in the case of the Sakai-Sugimoto model, ending on the flavor \D8-$\overline \D 8$-branes It would be interesting to understand the relations between these two proposals and the one based on the open string tachyon.  

\section*{Acknowledgments}

We would like to thank P. Argyres, S. Baharian, O. Bergman, and especially 
J. F. V\'azquez-Poritz for helpful discussions and comments . This work is supported by 
DOE grant DE-FG02-91ER40709. 

\appendix

\section{Free energies of possible bulk backgrounds}
Following \cite{asy0604} we use the notation adopted in \cite{ks0403} to compute the free energy of possible finite temperature $\Dq$-brane backgrounds for both compact and non-compact $x^q$-direction. We first consider the case of compact $x^q$ and as a warm-up  first do the computation for $\D6$-branes, then generalize it to  $\Dq$-branes. 

\subsection{Compact $x^q$}

\paragraph{$\D6$-branes} Consider the bosonic part of type IIA supergravity action (in Euclidean signature and) in string frame
\bea\label{D6sugraaction}
S &=& -(S_{\rm EH} + S_{\phi} + S_{\rm RR}+S_{\rm NS}), \nonumber\\
&=& -\int e^{-2\phi}\sqrt{g}({\cal R} + 4 \p\phi \p\phi ) + \frac{1}{2}\sum_q\int F_{(q+2)}
\w *F_{(q+2)}+ \frac{1}{2}  \int e^{-2\phi} H_{(3)}.
\w *H_{(3)}.\label{action}
\eea
Having the near horizon geometry of $\D6$-branes in mind, we choose the following (radial) ansatz for the metric and RR 7-form $C_{(7)}$ 
\begin{align}\label{ansatz}
l_s^{-2}ds^2 =& d\tau^2 + e^{2\lambda(\tau)} dx^2_{\|} +
e^{2\tilde{\lambda}(\tau)} dx_c^2 + e^{2\nu(\tau)} d\Omega_2^2,\\
C_{(7)} =& A\h{3pt} dx^1 \w ...\w dx^5 \w dx^{6-c}\w dx^c,
\end{align}
where $d\Omega_2^2$ is the line element of the unit two-sphere. Certainly the backgrounds we considered in this paper can all be put into the form of our ansatz (\ref{ansatz}). Here $\tau$ is the radial coordinate, $x_{\|}$ are 6 (out of 7)
worldvolume directions, $x_c$ plays the role of time in black $\D6$-brane geometry and $x^6$ in thermal geometry. From now on we set $\l_s=1$. It could be restored into our results by dimensional analysis. 

The equation of motion of $C_{(7)}$ has the solution 
\bea\label{RRsol}
\dot A e^{-6\lambda - \tilde{\lambda} + 2\nu} = \rm const,
\eea
where the dot represents the derivative with respect to $\tau$. Substituting (\ref{RRsol}) into (\ref{action}), the RR part of the action reads
\bea\label{RRcrudeact}
- S_{\rm RR} =  \int Q^2 e^{6\lambda + \tilde{\lambda} - 2\nu} d\tau,
\eea
where $Q^2$ is an integration constant. It is  more convenient to define a new radial coordinate $d\rho = - e^{\varphi} d\tau$, where
\bea\label{varphi}
\varphi = 2\phi - 6\lambda - \tilde\lambda - 2\nu, 
\eea
in terms of which (\ref{RRcrudeact}) reads
\begin{equation}\label{action RR}
- S_{\rm RR} =  -\int Q^2 e^{6\lambda + \tilde{\lambda} - 2\nu -
\varphi} d\rho.
\end{equation}
Note that $\sqrt{g} e^{-2\phi} d\tau = - e^{-2\varphi} d\rho$ by which the
dilaton action is easily computed to give
\begin{equation}\label{action RR}
- S_\phi =  \int 4\phi'^2 d\rho,
\end{equation}
where prime denotes the derivative with respect to  $\rho$ . With our metric ansatz (\ref{ansatz}), the Ricci scalar is
\begin{align}
{\cal R} = 2e^{-2\nu} - 42 \dot\lambda^2 - 2 \dot{\tilde\lambda}^2 - 6
\dot\nu^2 - 12 \dot\lambda \dot{\tilde\lambda} - 24
\dot\lambda\dot\nu - 4 \dot{\tilde\lambda} \dot\nu - 12 \ddot\lambda
- 2 \ddot{\tilde\lambda} - 4 \ddot\nu.
\end{align}
Thus, expressed in terms of the $\rho$ coordinate, the Einstein-Hilbert action becomes
\bea\label{simplifiedact}
- S_{\rm EH} &=& \int {\cal R} e^{-2\varphi} d\rho,\nonumber\\
 &=& \int \Big[2e^{-2\nu - 2\varphi} - 42 \lambda'^2 - 2
\tilde\lambda'^2 - 6 \nu'^2 - 12 \lambda' \tilde\lambda' -
24\lambda' \nu' - 4 \tilde\lambda' \nu' - 12 (\lambda'' +
\lambda'\varphi')\nonumber\\
&&\h{15pt} - 2 (\tilde\lambda'' + \tilde\lambda' \varphi') - 4
(\nu'' + \nu' \varphi')\Big] d\rho.
\eea
Note that the terms with double primes are total derivatives and are cancelled by adding the Gibbons-Hawking term to (\ref{simplifiedact}). Using (\ref{varphi}), $S_{\rm EH}$ and $S_\phi$ add up to give
the following simple expression 
\begin{equation}
- S_{\rm EH} - S_\phi = \int \Big{(} 2e^{-2\nu - 2\varphi}  -
6\lambda'^2 -  \tilde\lambda'^2 - 2 \nu'^2  + \varphi'^2 \Big{)}
d\rho.
\end{equation}
Putting everything together, the total action reads
\begin{equation}\label{final action}
S = \int \Big{(} 2e^{-2\nu - 2\varphi}  - 6\lambda'^2 -
\tilde\lambda'^2 - 2 \nu'^2  + \varphi'^2 - Q^2 e^{6\lambda +
\tilde{\lambda} - 2\nu}\Big{)} d\rho.
\end{equation}

The equations of motion are
\begin{align}
\label{lambda eq}\lambda'' =& \frac{Q^2}{2}e^{-2\phi + 12\lambda + 2\tilde\lambda},\\
\label{tillambda eq}\tilde\lambda'' =& \frac{Q^2}{2}e^{-2\phi + 12\lambda + 2\tilde\lambda},\\
\label{nu eq}\nu'' =& e^{-4\phi + 12\lambda + 2\tilde\lambda + 2\nu}
- \frac{Q^2}{2} e^{-2\phi + 12\lambda + 2\tilde\lambda},\\
\label{phi eq}\phi'' =& \frac{3Q^2}{2}e^{-2\phi + 12\lambda +
2\tilde\lambda},
\end{align}
where $\varphi$ has been replaced by $\phi$ using (\ref{varphi}). Defining $\Phi = -2\phi + 12\lambda + 2\tilde\lambda$, (\ref{lambda
eq}), (\ref{tillambda eq}) and (\ref{phi eq}) give
\bea
\Phi'' = 4Q^2 e^{\Phi},
\eea
with the following solution
\bea\label{rnd}
\Phi = - 2 \ln{\Big(\frac{\sqrt{2}Q}{C C_1}\sinh{C_1\rho}\Big)} - 2\ln{C}.
\eea
The last term in (\ref{rnd}) is there just for convenience. Going back to
(\ref{lambda eq}), (\ref{tillambda eq}) and (\ref{phi eq}) and solving
for $\lambda$, $\tilde\lambda$ and $\phi$, one obtains
\begin{align}
\lambda =& - \frac{1}{4} \ln{\Big(\frac{\sqrt{2}Q}{g_s
C_1}\sinh{C_1\rho}\Big)} +
C_2^\lambda \rho\label{lambda},\\
\tilde\lambda =& - \frac{1}{4} \ln{\Big(\frac{\sqrt{2}Q}{g_s
C_1}\sinh{C_1\rho}\Big)} + C_2^{\tilde\lambda} \rho\label{tillambda},\\
\phi =& - \frac{3}{4} \ln{\Big(\frac{\sqrt{2}Q}{g_s C_1}\sinh{C_1\rho}\Big)}
+ C_2^\phi \rho + \ln{g_s}\label{phi},
\end{align}
with $6 C_2^\lambda + C_2^{\tilde\lambda} -C_2^{\phi} = 0$. The
constants $C_1$, $C_2^\lambda$, $C_2^{\tilde\lambda}$ and
$C_2^{\phi}$ are to be determined. Defining $\Phi_0 = \Phi +
2\ln{g_s}$, the equation of motion for $\nu$ becomes
\bea
\nu'' = e^{\frac{\Phi_0}{4}- 2C_2^\phi \rho - 4 \ln{g_s} + 2\nu } -
\frac{1}{2}Q^2 e^{\Phi}.
\eea
If we take $2\nu = \frac{3}{4\Phi_0} + 2 C_2^\phi\rho + C_\nu$, the
above equation is satisfied if
$$C_\nu = \ln{2Q^2g_s^2}.$$
Thus, we obtain
\begin{equation}\label{nu}
\nu = - \frac{3}{4} \ln{\Big(\frac{\sqrt{2}Q}{g_s C_1}\sinh{C_1\rho}\Big)} +
C_2^\phi \rho + \ln{\sqrt{2}Qg_s}.
\end{equation}\\

In what follows we show that the near horizon geometry of non-extremal $\D6$-branes satisfies (\ref{lambda}), (\ref{tillambda}), (\ref{phi}). This geometry takes the form
\begin{align}
ds^2 =& f^{-1/2}\Big{(}dx^2_{\|} + h dx_c^2 \Big{)} + f^{1/2}
\Big{(}\frac{1}{h}du^2+ u^2 d\Omega^2_2,
\Big{)}\label{NH solu:metric}\\
e^{-2\phi} =& g_s^{-2} f^{3/2}\label{NH solu:dilaton},
\end{align}
with
\bea
f = \frac{g_s N_c}{2l_s u}, \qquad {\rm and} \qquad h = 1 - \frac{u_c}{u},
\eea
where $u_c$ is either $u_T$ or $u_{\rm KK}$ (or zero for the geometry given in (\ref{compactmetric3})).
The $u$
coordinate is not the same as $\rho$. Their relation can
be inferred from (\ref{lambda}) and (\ref{tillambda}), and is given by
\bea
e^{2(\tilde\lambda - \lambda)} = e^{2\rho (C_2^{\tilde\lambda} - C_2^\lambda)} = h,
\eea
yielding
\bea\label{rho to u}
\rho &=& \frac{1}{2\Delta C_2} \ln{h},\nonumber \\
d\rho &=& \frac{u_c}{2u^2\Delta C_2} \frac{1}{h}du,
\eea
where $\Delta C_2 = C_2^{\tilde\lambda} - C_2^\lambda$. Computing $e^{2\lambda}$ from (\ref{lambda}), we have
\bea\label{elanda}
e^{2\lambda} &=&
\frac{e^{2C_2^\lambda\rho}}{\sqrt{\Big(\frac{\sqrt{2}Q}{g_s
C_1}\Big)\sinh{C_1\rho}}} ,\nonumber\\
&=& \sqrt{\frac{\sqrt{2}C_1 g_s}{Q}}
\frac{h^{\frac{C_2^\lambda}{\Delta
C_2}}}{\sqrt{h^{\frac{C_1}{2\Delta C_2}} - h^{\frac{-C_1}{2\Delta
C_2}}} },\nonumber\\
&=& \sqrt{\frac{\sqrt{2}C_1 g_s}{Q}}
\frac{h^{\frac{C_2^\lambda}{\Delta C_2} - \frac{C_1}{4\Delta
C_2}}}{\sqrt{1- h^{\frac{-C_1}{\Delta C_2}}} }.
\eea
To match $e^{2\lambda}$ in (\ref{elanda}) with $f^{1/2}$, one has to impose $-C_1
= \Delta C_2$ and $C_1 = 4 C_2^\lambda$, or
\bea\label{C relation 1}
C_1 = 4 C_2^\lambda; \h{15pt} C_2^{\tilde\lambda} = -3 C_2\lambda.
\eea
Thus, we have
\bea\label{clambda}
e^{2\lambda} =& \sqrt{\frac{\sqrt{2}C_1 g_s}{Q}\frac{u}{u_c}}
\equiv \Big(\frac{2l_s u}{g_s N}\Big)^{1/2} \qquad
\Rightarrow \qquad C_2^\lambda = \frac{Qu_c l_s}{2\sqrt{2} g_s^2 N}.
\eea
Taking $C_2^\phi = 3 C_2^\lambda$, we can match the dilaton with the dilaton of the $\D6$-brane solution (\ref{NH solu:dilaton})
\bea
e^{2\phi} = e^{6\lambda + 2\ln{g_s}} = g^2_s \Big(\frac{2l_s u}{g_s N_c}\Big)^{3/2}.
\eea
Furthermore, from (\ref{nu}), we have 
\bea\label{Nequat}
e^{2\nu} = 2Q^2 g_s^2 e^{6\lambda} =2Q^2g_s^2 \Big(\frac{2l_s
u}{g_sN_c}\Big)^{3/2} \equiv \Big(\frac{g_s N}{2l_s u}\Big)^{1/2} u^2 \qquad 
\Rightarrow \qquad N =2\sqrt{2} Ql_s
\eea
The final check is the $g_{uu}$ component of the metric. To do that, we first need
to compute $e^{-2\varphi}$ which turns out to be
\bea
e^{-2\varphi} &=& e^{-4\phi + 12\lambda + 2\tilde\lambda + 6 \nu},\nonumber\\
&=&e^{14\lambda - 8 C_2^\lambda\rho + 4 \ln{(\sqrt{2}Qg_s)} - 4
\ln{g_s}},\nonumber\\
&=& \Big(\frac{2l_s u}{g_s N}\Big)^{7/2} h (\sqrt{2}Q)^4.
\eea
Hence,
\bea\label{taurho}
d\tau^2 &=& e^{-2\varphi} d\rho^2,\nonumber\\ 
&=&\Big(\frac{2l_s u}{g_sN}\Big)^{7/2} h
(\sqrt{2}Q)^4 \frac{u_c^2}{4u^4 \Big(16 (C_2^{\lambda})^2\Big)} \frac{1}{h^2},\nonumber\\
&=& \Big(\frac{g_s N_c}{2 l_s u }\Big)^{1/2} \frac{1}{h},
\eea
where we have used (\ref{clambda}) and (\ref{Nequat}) to simplify the
expression. Thus, the near horizon geometry of $N_c$ $\D6$-branes is indeed a solution to type IIA supergravity equations of motion.

It is now straightforward to plug (\ref{lambda}), (\ref{tillambda}), (\ref{phi}) and (\ref{nu}) into (\ref{final action}) to obtain the on-shell action (free energy). Changing the coordinate from $\rho$ coordinate to $u$ and being
careful about the fact that $d\rho/du$ is negative, we obtain 
\bea\label{D6freeenergy}
S&=& - V_9 \int_{u_c}^\infty \Big{\{}  \Big{(}- 6\lambda'^2 -
\tilde\lambda'^2 - 2 \nu'^2  + \varphi'^2\Big{)} \frac{du}{d\rho}+
\Big{(}2e^{-2\nu - 2\varphi} - Q^2 e^{6\lambda + \tilde{\lambda} -
2\nu}\Big{)}\frac{d\rho}{du} \Big{\}} du\\
&=& \frac{3}{g_s^2}V_9 \int_{u_c}^{\Lambda}  du=\frac{3}{g_s^2}V_9 (\Lambda-u_c).
\eea
where in the first line prime denotes the $u$-derivative, and $V_9$ is the volume (in units of string length which we set equal to one) of space transverse to $u$: $V_9=\beta  {\rm Vol}({\rm S}^1_{x^q}) {\rm Vol}({\rm S}^{2}){\rm Vol}(\mathbb{R}^{5})$. The free energy is infinite. That is why we put a cutoff $u=\Lambda$ in the second line to regulate the free energy. The cutoff drops out when it comes to comparing the free energies of different solutions. 

It is clear from (\ref{D6freeenergy}) that a $D6$-brane solution of the type (\ref{compactmetric3}) has always more free energy than a black $\D6$-brane (\ref{compactmetric2}) or a thermal $\D6$-brane (\ref{compactmetric1}). Comparing the free energies of  the black and thermal $\D6$-brane geometries, we obtain 
\bea
S_{\rm thermal~\D6} - S_{\rm black~\D6-brane}&=& \frac{3}{g_s^2}V_9 
(u_T-u_{\rm KK}),\\
&=&\frac{3V_9}{16\pi^2 g_s^2 R_{6+1}} (\beta^2-\beta_c^2),
\eea
where in the second line we have used (\ref{xqperiod}) and (\ref{timeperiod}), and 
\bea
\beta_c=2\pi R_c= \frac{4\pi}{3}\sqrt{R_{6+1}u_{\rm KK}}.
\eea
Thus, there is a phase transition at $\beta=2\pi R_c$, where the thermal $\D6$-brane geometry dominates for $\beta<2\pi R_c$ whereas for $\beta>2\pi R_c$ the black brane geometry dominates. 

\paragraph{$\Dq$-branes: general analysis} Having gone through the analysis for the $\D6$-brane, calculating the free energy of non-extremal $\Dq$-branes is straightforward. The relevant parts of the supergravity action (in string frame) are
\bea
S &=& -(S_{\rm EH} + S_{\phi} + S_{\rm RR}+S_{\rm NS}), \nonumber\\
&=& -\int e^{-2\phi}\sqrt{g}({\cal R} + 4 \p\phi \p\phi ) + \frac{1}{2}\sum_q\int F_{(q+2)}
\w *F_{(q+2)}+ \frac{1}{2}  \int e^{-2\phi}H_{(3)}
\w *H_{(3)}.\label{action-p}
\eea
We are interested in backgrounds with $H_{(3)}$ not turned on, and choose the following ansatz for the metric and the RR $q$-form $C_{q+1}$
\bea\label{ansatz-p}
ds^2 &=& d\tau^2 + e^{2\lambda(\tau)} dx^2_{\|} +
e^{2\tilde{\lambda}(\tau)} dx_c^2 + e^{2\nu(\tau)} d\Omega_k^2,\\
C_{(q+1)} &=& A\h{3pt} dx^1 \w ...\w dx^{q-1}\w dx^{q-c} \w dx^c,
\eea
where $k = 8 - q$, and  $d\Omega_k^2$ is the line element of unit $k$-sphere. Again, we have set $l_s=1$. The equation of motion for $C_{(q+1)}$ has the solution
\bea
\dot A e^{-q\lambda - \tilde{\lambda} + k\nu} = \rm const.
\eea
The RR action then becomes
\bea
- S_{\rm RR} =  \int Q^2 e^{q\lambda + \tilde{\lambda} - k\nu} d\tau,
\eea
where, again, we denoted the integration constant by $Q^2$. For convenience, we change the variable from $\tau$ to $\rho$ defined by $d\rho = - e^{\varphi} d\tau$, where
\bea\label{varphi-p}
\varphi = 2\phi - q\lambda - \tilde\lambda - k\nu.
\eea
Then
\begin{equation}\label{action RR-p}
- S_{\rm RR} =  -\int Q^2 e^{q\lambda + \tilde{\lambda} - k\nu -
\varphi} d\rho.
\end{equation}

Following the same argument as we did for $q = 6$, it is not hard to see
that the action (\ref{action-p}) takes the form
\bea\label{final action-p}
S = \int \Big{(} k(k-1)e^{-2\nu - 2\varphi}  - q\lambda'^2 -
\tilde\lambda'^2 - k \nu'^2  + \varphi'^2 - Q^2 e^{q\lambda +
\tilde{\lambda} - k\nu}\Big{)} d\rho.
\eea

Having put the action in the above form, the equations of motion are now in order
\bea
\label{lambda eq-p}\lambda'' &=& \frac{Q^2}{2}e^{-2\phi + 2q\lambda + 2\tilde\lambda}\\
\label{tillambda eq-p}\tilde\lambda'' &=& \frac{Q^2}{2}e^{-2\phi + 2q\lambda + 2\tilde\lambda}\\
\label{nu eq-p}\nu'' &=& (k-1)e^{-2\nu - 2\varphi}
- \frac{Q^2}{2} e^{-2\phi + 2q\lambda + 2\tilde\lambda}\\
\label{phi eq-p}\phi'' &=& \frac{(5-k)Q^2}{2}e^{-2\phi + 2q\lambda +
2\tilde\lambda},
\eea
Define $\Phi = -2\phi + 2q\lambda + 2\tilde\lambda$, (\ref{lambda
eq-p}), then (\ref{tillambda eq-p}) and (\ref{phi eq-p}) result in
\bea
\Phi'' = 4Q^2 e^{\Phi},
\eea
which yields the following solution
\begin{equation}
\Phi = - 2 \ln{(\frac{\sqrt{2}Q}{C C_1}\sinh{C_1\rho})} - 2\ln{C}.
\end{equation}
The last term is there just for convenience. Solving for $\lambda$, $\tilde\lambda$ and $\phi$ in 
(\ref{lambda eq-p}), (\ref{tillambda eq-p}) and (\ref{phi eq-p}), we obtain
\bea
\lambda &=& - \frac{1}{4} \ln{\Big(\frac{\sqrt{2}Q}{g_s
C_1}\sinh{C_1\rho}\Big)} +
C_2^\lambda \rho,\label{lambda-p}\\
\tilde\lambda&=& - \frac{1}{4} \ln{\Big(\frac{\sqrt{2}Q}{g_s
C_1}\sinh{C_1\rho}\Big)} + C_2^{\tilde\lambda} \rho, \label{tillambda-p}\\
\phi &=& - \frac{5-k}{4} \ln{\Big(\frac{\sqrt{2}Q}{g_s
C_1}\sinh{C_1\rho}\Big)} + C_2^\phi \rho + \ln{g_s}\label{phi-p},
\eea
with $q C_2^\lambda + C_2^{\tilde\lambda} - C_2^{\phi} = 0$. The
constants $C_1$, $C_2^\lambda$, $C_2^{\tilde\lambda}$ and
$C_2^{\phi}$ are to be determined. Define $\Phi_0 = \Phi +
2\ln{g_s}$, then the equation of motion for $\nu$ becomes
\begin{equation}
\nu'' = (k-1)e^{\frac{7-q}{4}\Phi_0- 2C_2^\phi \rho - 4 \ln{g_s} +
2(k-1)\nu } - \frac{1}{2}Q^2 e^{\Phi}.
\end{equation}
If we take $2(k-1)\nu = \frac{q-3}{4}\Phi_0 + 2 C_2^\phi\rho +
C_\nu$, then the above equation is satisfied if
\bea
C_\nu = \ln{\frac{2Q^2g_s^2}{(k-1)^2}}.
\eea
Thus, we obtain
\bea\label{nu-p}
\nu = - \frac{5-k}{4(k-1)} \ln{\Big(\frac{\sqrt{2}Q}{g_s
C_1}\sinh{C_1\rho}\Big)} + \frac{C_2^\phi \rho}{k-1} +
\frac{1}{k-1}\ln{\frac{\sqrt{2}Qg_s}{k-1}}.
\eea

We now check whether the near horizon geometry of non-extremal $\Dq$-branes satisfies (\ref{lambda-p}), (\ref{tillambda-p}), (\ref{phi-p}) and (\ref{nu-p}). This geometry takes the form
\bea
ds^2 &=& f_q^{-1/2}\Big{(}dx^2_{\|} + h dx_c^2 \Big{)} + f_q^{1/2}
\Big{(}\frac{1}{h}du^2+ u^2 d\Omega^2_k
\Big{)},\label{NH solu:metric-p}\\
e^{-2\phi} &=& g_s^{-2} f_q^{(q-3)/2}\label{NH solu:dilaton-p},
\eea
where $f _q= \frac{d_q' g_s N_c}{(l_s u)^{7-q}}$, $d_p' =
(2\sqrt{\pi})^{5-q} \Gamma(\frac{7-q}{2})$ and $h = 1 -(u_c/u)^{7-q}$.The relation between $u$ and $pho$ can is inferred from
(\ref{lambda-p}) and (\ref{tillambda-p})
\bea
e^{2(\tilde\lambda - \lambda)} = e^{2\rho (C_2^{\tilde\lambda} - C_2^\lambda)} = h,
\eea
which gives
\bea\label{rho to u-p}
\rho =& \frac{1}{2\Delta C_2} \ln{h} \qquad \Rightarrow \qquad d\rho = \frac{(7-q)u^{7-q}_c u^{q-8}}{2\Delta C_2} \frac{1}{h}du,
\eea
where $\Delta C_2 = C_2^{\tilde\lambda} - C_2^\lambda$. Calculating $e^{2\lambda}$ from (\ref{lambda-p}), we have
\bea\label{elambdap}
e^{2\lambda} &=&\frac{e^{2C_2^\lambda\rho}}{\sqrt{\Big(\frac{\sqrt{2}Q}{g_s
C_1}\Big)\sinh{C_1\rho}}}\nonumber\\
& =& \sqrt{\frac{\sqrt{2}C_1 g_s}{Q}}\frac{h^{\frac{C_2^\lambda}{\Delta
C_2}}}{\sqrt{h^{\frac{C_1}{2\Delta C_2}} - h^{\frac{-C_1}{2\Delta
C_2}}} } \nonumber\\
&=& \sqrt{\frac{\sqrt{2}C_1 g_s}{Q}}
\frac{h^{\frac{C_2^\lambda}{\Delta C_2} - \frac{C_1}{4\Delta
C_2}}}{\sqrt{1- h^{\frac{-C_1}{\Delta C_2}}} }.
\eea
The only way to make $e^{2\lambda}$ in (\ref{elambdap}) to be equal to $f_q^{1/2}$ is by imposing
$-C_1 = \Delta C_2$ and $C_1 = 4 C_2^\lambda$, or
\bea\label{C relation 1-p}
C_1 = 4 C_2^\lambda; \h{15pt} C_2^{\tilde\lambda} = -3 C_2\lambda.
\eea
Thus, we have
\bea\label{C2lambda-p}
e^{2\lambda} = \sqrt{\frac{\sqrt{2}C_1 g_s}{Q}}
\Big{(}\frac{u}{u_c}\Big{)}^{\frac{7-q}{2}}
\equiv f_q^{-1/2}\qquad \Rightarrow\qquad  C_2^\lambda = \frac{Q(u_c l_s)^{7-q}}{4\sqrt{2} g_s^2 N_cd_q'}.
\eea
Also, the dilaton can be matched to (\ref{NH solu:dilaton-p}) as long as we take $C_2^\phi = (5-k) C_2^\lambda$ resulting in
\bea
e^{2\phi} = e^{2\lambda(5-k) + 2\ln{g_s}} = g^2_s f_q^{-\frac{q-3}{2}}.
\eea
Furthermore, from (\ref{nu-p}), we have
\bea\label{N-p}
e^{2\nu} = \Big{(}\frac{\sqrt{2}Qg_s}{k-1}
\Big{)}^{\frac{2}{k-1}} e^{2\lambda\frac{5-k}{k-1}} \equiv f_q^{1/2} u^2\qquad
\Rightarrow \qquad N_c =& \frac{\sqrt{2} Ql_s^{k-1}}{(k-1)d_q'}.
\eea
The final check is the $g_{uu}$ component. To do that, we first need
to compute $e^{-2\varphi}$ which gives 
\bea
e^{-2\varphi} &=& e^{-4\phi + 2q\lambda + 2\tilde\lambda + 2k \nu}\nonumber\\
&=&e^{\frac{6k+2}{k-1}\lambda - 8 C_2^\lambda\rho +\frac{2k}{k-1}
\ln{(\frac{\sqrt{2}Qg_s}{k-1})} - 4 \ln{g_s}}.
\eea
Hence
\bea
d\tau^2 &=&  e^{-2\varphi} d\rho^2\nonumber\\
&=&\frac{(7-q)^2u^{2(7-q)}_c u^{2(q-8)}}{64 (C^{\lambda}_2)^2}
\frac{1}{h^2} e^{\frac{6k+2}{k-1}\lambda - 8 C_2^\lambda\rho +
\frac{2k}{k-1}\ln{(\frac{\sqrt{2}Qg_s}{k-1})} - 4 \ln{g_s}} du^2\nonumber \\
&=& f_q^{1/2} \frac{1}{h} du^2,
\eea
where we have used (\ref{C2lambda-p}) and (\ref{N-p}) to simplify
the expression. Therefore, the exact solution we just found indeed
corresponds to the near horizon geometry of non-extremal $N_c$ $\Dq$-branes.

With (\ref{lambda-p}), (\ref{tillambda-p}), (\ref{phi-p}) and
(\ref{nu-p}) at hand, it is now straightforward to plug them into (\ref{final
action-p}) to get the on-shell action. Changing back to the $u$ coordinate and being careful about the fact that $d\rho/du$ is
negative, we obtain (here prime denotes $u$
derivative.)
\bea\label{freeenergyDq}
S&=& - V_9 \int_{u_c}^\infty \Big{\{}  \Big{(}-
q\lambda'^2 - \tilde\lambda'^2 - k \nu'^2  + \varphi'^2\Big{)}
\frac{du}{d\rho}+ \Big{(}k(k-1)e^{-2\nu - 2\varphi} - Q^2
e^{q\lambda + \tilde{\lambda} -
k\nu}\Big{)}\frac{d\rho}{du} \Big{\}}du\nonumber\\
&=& \frac{9-q}{g_s^2} V_9 \int_{u_c}^\Lambda (7-q)~u^{6-q} du\nonumber\\
&=& \frac{9-q}{g_s^2} V_9 \Big{(}\Lambda^{7-q} - u_c^{7-q} \Big{)},
\eea
where $\Lambda$ is the cutoff, and $V_9=\beta R_c {\rm Vol}({\rm S}^{8-q}){\rm Vol}(\mathbb{R}^{q-1})$ (in units where string length $l_s=1$). For $q\leq4$, it is not hard to see from (\ref{freeenergyDq}) that the geometry given in (\ref{compactmetric3}) for which $u_c=0$ has more free energy than the geometries given in (\ref{compactmetric1}) and (\ref{compactmetric2}). The difference in free energies of the thermal and black $\Dq$-brane geometries is given by
\bea
S_{\rm thermal} - S_{\rm black~brane}= \frac{9-q}{g_s^2} V_9 \Big{(}u_T^{7-q} - u_{\rm KK}^{7-q} \Big{)},
\eea
which using (\ref{xqperiod}) and (\ref{timeperiod}) can be equivalently expresses as
\bea
S_{\rm thermal}-S_{\rm black~brane}=\frac{9-q}{g_s^2} V_9 \left(\frac{4\pi}{7-q}R_{q+1}^{\frac{1}{2}(7-q)} \right)^{2\frac{7-q}{5-q}}\left(\beta^{2\frac{q-7}{5-q}}-\beta_c^{2\frac{q-7}{5-q}}\right).
\eea
Thus, for $q\leq4$ there is a phase transition at $\beta=2\pi R_c$. For $\beta>2\pi R_c$ the thermal geometry (\ref{compactmetric1}) dominates whereas for $\beta<2\pi R_c$ it is the black brane geometry
(\ref{compactmetric2}) which dominates the Euclidean path integral. The case of $\D5$-branes and any associated phase transition has been discussed in section 5.

\subsection{Non-compact $x^q$}
When $x^q$ is not compact, and $q\neq5$, there are only two geometries with the same asymptotic boundary condition: the thermal geometry (\ref{Dqthermal}) and the black brane geometry (\ref{Dq metric}) . The difference in free energies of the two geometries is obtained by first setting $u_c=0$ in (\ref{freeenergyDq}) for the thermal and  $u_c=u_T$ for the black brane geometry, then subtracting the results
\bea\label{dif}
S_{\rm thermal}-S_{\rm black~brane}=\frac{9-q}{g_s^2} V_9u_T^{7-q},
\eea
where $V_9={\rm Vol}({\rm S}^{8-q}){\rm Vol}(\mathbb{R}^{q}){\rm Vol}({\rm S}^1_{\beta})$ in units where $l_s=1$.  The difference in free energies (\ref{difffree}) is positive indicating that independent of temperature the black brane geometry is always dominant and determines the vacuum.  The case of $\D5$-branes has been discussed in section 2.


\begin{thebibliography}{99}

\bibitem{ss0412}
T.~Sakai and S.~Sugimoto, ``Low energy hadron physics in holographic QCD,''
Prog.\ Theor.\ Phys.\  {\bf 113}, 843 (2005), [hep-th/0412141].

\bibitem{m9711} 
J. M. Maldacena, ``The large N limit of superconformal field theories and 
supergravity,'' Adv. Theor. Math. Phys. {\bf 2} (1998) 231; Int. J. Theor. 
Phys. {\bf 38} (1999) 1113, [hep-th/9711200].

\bibitem{gkp9802} 
S. S. Gubser, I. R. Klebanov and A. M. Polyakov, ``Gauge theory correlators 
from non-critical string theory,'' Phys.\ Lett.\ B {\bf 428}, 105 (1998), 
[hep-th/9802109].

\bibitem{w9802} 
E.~Witten, ``Anti-de Sitter space and holography,'' Adv.\ Theor.\ Math.\ 
Phys.\  {\bf 2}, 253 (1998), [hep-th/9802150].

\bibitem{agmoo} 
O. Aharony, S. S. Gubser, J. Maldacena, H. Ooguri and
Y. Oz, ``Large $N$ field theories, string theory and gravity,'' Phys. 
Rept. {\bf 323} (2000) 183, [hep-th/9905111].

\bibitem{w9803}
E.~Witten, ``Anti-de Sitter space, thermal phase transition, and confinement in  gauge theories,'' Adv.\ Theor.\ Math.\ Phys.\  {\bf 2}, 505 (1998), [hep-th/9803131].

\bibitem{bisy9803}
A.~Brandhuber, N.~Itzhaki, J.~Sonnenschein and S.~Yankielowicz, 
``Wilson loops in the large N limit at finite temperature,'' Phys.\ Lett.\  B {\bf 434}, 36 (1998), [hep-th/9803137].

\bibitem{ahjk0604}
E.~Antonyan, J.~A.~Harvey, S.~Jensen and D.~Kutasov, ``NJL and QCD from string
theory,'' [hep-th/0604017].

\bibitem{njl1961}
Y.~Nambu and G.~Jona-Lasinio, ``Dynamical model of elementary particles based 
on an analogy with superconductivity. I,'' Phys.\ Rev.\  {\bf 122}, 345 (1961).

\bibitem{asy0604}
O.~Aharony, J.~Sonnenschein and S.~Yankielowicz, ``A holographic model of
deconfinement and chiral symmetry restoration,'' Annals Phys.\  {\bf 322},
1420 (2007), [hep-th/0604161].

\bibitem{ps0604}
A.~Parnachev and D.~A.~Sahakyan,
``Chiral phase transition from string theory,'' Phys.\ Rev.\ Lett.\  {\bf 97},
111601 (2006), [hep-th/0604173].

\bibitem{ahk0608}
E.~Antonyan, J.~A.~Harvey and D.~Kutasov, ``Chiral symmetry breaking from
intersecting D-branes,''
Nucl.\ Phys.\  B {\bf 784}, 1 (2007), [hep-th/0608177].

\bibitem{imsy9802}
N.~Itzhaki, J.~M.~Maldacena, J.~Sonnenschein and S.~Yankielowicz,
``Supergravity and the large N limit of theories with sixteen supercharges,''
Phys.\ Rev.\  D {\bf 58}, 046004 (1998), [hep-th/9802042].

\bibitem{a9911}
O.~Aharony, ``A brief review of 'little string theories','' 
Class.\ Quant.\ Grav.\  {\bf 17}, 929 (2000), [hep-th/9911147].

\bibitem{kmmw0311}
M.~Kruczenski, D.~Mateos, R.~C.~Myers and D.~J.~Winters, ``Towards a holographic
dual of large-N(c) QCD,'' JHEP {\bf 0405}, 041 (2004), [hep-th/0311270].

\bibitem{hnt0710}
N.~Horigome, M.~Nishimura and Y.~Tanii, ``Chiral Symmetry Breaking in Brane Models,''
arXiv:0710.4900 [hep-th].

\bibitem{gxz0605}
Y.~h.~Gao, W.~s.~Xu and D.~f.~Zeng, ``NGN, QCD(2) and chiral phase transition
from string theory,'' JHEP {\bf 0608}, 018 (2006), [hep-th/0605138].

\bibitem{gn1974}
D.~J.~Gross and A.~Neveu, ``Dynamical Symmetry Breaking In Asymptotically Free 
Field Theories,'' Phys.\ Rev.\  D {\bf 10}, 3235 (1974).

\bibitem{fgmp0704}
J.~J.~Friess, S.~S.~Gubser, G.~Michalogiorgakis and S.~S.~Pufu, ``Stability of 
strings binding heavy-quark mesons,'' JHEP {\bf 0704}, 079 (2007), [hep-th/0609137].

\bibitem{ht0608}
N.~Horigome and Y.~Tanii, ``Holographic chiral phase transition with chemical potential,'' JHEP {\bf 0701}, 072 (2007), [hep-th/0608198].

\bibitem{bll0708}
O.~Bergman, G.~Lifschytz and M.~Lippert, ``Holographic Nuclear Physics,'' JHEP {\bf 0711}, 056 (2007),  arXiv:0708.0326 [hep-th].

\bibitem{dgks0708}
J.~L.~Davis, M.~Gutperle, P.~Kraus and I.~Sachs, ``Stringy NJL and Gross-Neveu 
models at finite density and temperature,'' JHEP {\bf 0710}, 049 (2007), arXiv:0708.0589 
[hep-th].

\bibitem{rsvw0708}
M.~Rozali, H.~H.~Shieh, M.~Van Raamsdonk and J.~Wu, ``Cold Nuclear Matter In 
Holographic QCD,'' JHEP {\bf 0801}, 053 (2008), arXiv:0708.1322 [hep-th].


\bibitem{bll0802}
O.~Bergman, G.~Lifschytz and M.~Lippert, ``Response of Holographic QCD to 
Electric and Magnetic Fields,'' arXiv:0802.3720 [hep-th].

\bibitem{jk0803}
C.~V.~Johnson and A.~Kundu, ``External Fields and Chiral Symmetry Breaking 
in the Sakai-Sugimoto Model,'' arXiv:0803.0038 [hep-th].

\bibitem{ksz0803}
K.~Y.~Kim, S.~J.~Sin and I.~Zahed, ``Dense and Hot Holographic QCD: Finite 
Baryonic E Field,'' arXiv:0803.0318 [hep-th].

\bibitem{g0411}
M.~R.~Garousi, ``D-brane anti-D-brane effective action and brane interaction in open
string channel,'' JHEP {\bf 0501}, 029 (2005), [hep-th/0411222].

\bibitem{ckp0702}
R.~Casero, E.~Kiritsis and A.~Paredes, ``Chiral symmetry breaking as open string
tachyon condensation,'' Nucl.\ Phys.\  B {\bf 787}, 98 (2007), [hep-th/0702155].

\bibitem{bss0708}
O.~Bergman, S.~Seki and J.~Sonnenschein, ``Quark mass and condensate in HQCD,''
arXiv:0708.2839 [hep-th].


\bibitem{dn0708}
A.~Dhar and P.~Nag, ``Sakai-Sugimoto model, Tachyon Condensation and Chiral
symmetry Breaking,'' arXiv:0708.3233 [hep-th].

\bibitem{eln}
M. Edalati, R. G. Leigh and N. Nguyen, work in progress.

\bibitem{dgk0109} 
U.~H.~Danielsson, A.~Guijosa and M.~Kruczenski, ``Brane-antibrane systems at 
finite temperature and the entropy of black branes,'' JHEP {\bf 0109}, 011 
(2001), [hep-th/0106201].

\bibitem{h0212}
K.~Hotta, ``Brane-antibrane systems at finite temperature and phase transition 
near the Hagedorn temperature,'' JHEP {\bf 0212}, 072 (2002), [hep-th/0212063].

\bibitem{ct0802}
V.~Calo and S.~Thomas, ``Phase Transitions in Separated $D_{p-1}$ and anti-$D_{p-1}$ 
Branes at Finite Temperature,'' arXiv:0802.2453 [hep-th].

\bibitem{ak0803}
O.~Aharony and D.~Kutasov, ``Holographic Duals of Long Open Strings,'' Phys.\ Rev.\  D {\bf 78}, 026005 (2008), arXiv:0803.3547 [hep-th].

\bibitem{mms0807}
R.~McNees, R.~C.~Myers and A.~Sinha, ``On quark masses in holographic QCD,'' 
JHEP {\bf 0811}, 056 (2008), arXiv:0807.5127 [hep-th].

\bibitem{aelv0811}
P.~C.~Argyres, M.~Edalati, R.~G.~Leigh and J.~F.~Vazquez-Poritz, ``Open Wilson Lines and Chiral Condensates in Thermal Holographic QCD,'' arXiv:0811.4617 [hep-th].

\bibitem{hhly0803}
K.~Hashimoto, T.~Hirayama, F.~L.~Lin and H.~U.~Yee, ``Quark Mass Deformation of Holographic Massless QCD,'' JHEP {\bf 0807}, 089 (2008), arXiv:0803.4192 [hep-th].

\bibitem{ks0403}
S.~Kuperstein and J.~Sonnenschein, ``Non-critical supergravity ($d > 1$) and holography,'' JHEP {\bf 0407}, 049 (2004), [hep-th/0403254].

\end{thebibliography}
\end{document}